\documentclass[12pt]{article}

\usepackage{color}
\usepackage[english]{babel}
\usepackage{amsmath, amsfonts, amssymb}
\usepackage{mathrsfs}
\usepackage{graphicx}
\usepackage[colorlinks=true,linkcolor=blue]{hyperref}
\usepackage{cite}
\usepackage{cleveref}
\usepackage{physics}
\usepackage{fontawesome5}
\usepackage[dvipsnames]{xcolor}
\usepackage{color}
\usepackage{upgreek}
\usepackage{caption}
\usepackage{feynmp}
\usepackage{tikz}
\usepackage{relsize}
\newcommand{\github}[1]{%
   \href{#1}{\faGithubSquare}%
}
\def\ba{\begin{eqnarray}}
\def\ea{\end{eqnarray}}
\def\be{\begin{equation}}
\def\ee{\end{equation}}
\def\nn{\nonumber}
\def\d{\partial}

\renewcommand{\l}{\lambda}
\renewcommand{\L}{\Lambda}
\renewcommand{\O}{\Omega}
\renewcommand{\o}{\omega}
\renewcommand{\k}{\kappa}
\renewcommand{\b}{\beta}
\renewcommand{\H}{\mathcal{H}}

\newcommand{\g}{\gamma}
\newcommand{\vk}{\varkappa}

\newcommand{\G}{\mathcal{G}}

\newcommand{\vf}{\varphi}
\newcommand{\gc}{\mathrm{g}}
\newcommand{\diff}{\mathrm{d}}
\newcommand{\e}{{\rm e}}
\newcommand{\bb}{\vf_{\mathrm{b}}}
\newcommand{\C}{\mathscr{C}}

\renewcommand{\Re}{\mathop{\rm Re}\nolimits}
\renewcommand{\Im}{\mathop{\rm Im}\nolimits}
\newcommand{\ch}{\mathop{\rm ch}\nolimits}
\newcommand{\sh}{\mathop{\rm sh}\nolimits}
\newcommand{\arctg}{\mathop{\rm arctg}\nolimits}

\renewcommand{\b}{\beta}
\renewcommand{\g}{\gamma}

\newcommand{\eff}{{\rm eff}}

\newcommand{\near}{\rm near}
\newcommand{\far}{\rm far}
\newcommand{\bl}{\bar{\lambda}}
\newcommand{\dl}{\delta_{\lambda}}
\newcommand{\Bl}{\bar{\Lambda}}
\newcommand{\Dl}{\delta_{\Lambda}}

\newcommand\bseq{\begin{subequations}}
\newcommand\eseq{\end{subequations}}

\usepackage{etoolbox}
\preto\subequations{\ifhmode\unskip\fi}

\allowdisplaybreaks[4]

\begin{document}

\title{
False vacuum decay catalyzed by black hole\\ in a heat bath
}

\author{
Bowen Hu$^{1,2,3}$\thanks{hubowen24@mails.ucas.ac.cn}~,~ 
Kohei Kamada$^{1,4,5}$\thanks{kohei.kamada@ucas.ac.cn}~,~ 
Andrey Shkerin$^{6}$\thanks{ashkerin@perimeterinstitute.ca}
\\[2mm]
{\small\it $^1$School of Fundamental Physics and Mathematical Sciences, Hangzhou Institute for Advanced Study,} \\
{\small\it University of Chinese Academy of Sciences (HIAS-UCAS), 310024 Hangzhou, China} \\
{\small\it $^2$Institute of Theoretical Physics, Chinese Academy of Sciences, Beijing 100190, China}\\
{\small \it $^3$University of Chinese Academy of Sciences, Beijing 100049, China} \\
{\small \it $^4$International Centre for Theoretical Physics Asia-Pacific (ICTP-AP), Hangzhou/Beijing, China} \\
{\small \it $^5$Research Center for the Early Universe, The University of Tokyo, Bunkyo-ku, Tokyo 113-0033, Japan} \\
{\small\it $^6$Perimeter Institute for Theoretical Physics,}
{\small\it 31 Caroline St N, Waterloo, ON N2L 2Y5, Canada}
}

\date{}

\maketitle

\begin{abstract}

We study false vacuum decay catalyzed by black holes. We consider a scalar field model with unstable potential in the background of a dilaton black hole in two dimensions. The model reproduces many features of the Schwarzschild black hole background in four dimensions, including the centrifugal barrier for linearized field perturbations. 
We study decays from the non-equilibrium state describing the evaporating black hole immersed in the thermal bath with a different temperature. 
We analytically construct the tunneling solution relevant at small field excitations and evaluate the decay suppression. We show how they reduce to those for the Hartle-Hawking (equilibrium) and Unruh states in the corresponding limits. 
For large field excitations the decay proceeds via stochastic activation;
we find the relevant non-thermal sphaleron configuration in a certain region of parameters of the model and construct the semiclassical solution describing tunneling onto this sphaleron.
Our results provide insights into the vacuum decay induced by small primordial black holes in the radiation-dominated era of the universe.

\end{abstract}

\newpage
{\hypersetup{hidelinks}
\tableofcontents
}

\section{Introduction}
\label{sec:intro}

False vacuum decay in quantum field theory is an important area of research with many applications to high-energy physics phenomenology and cosmology.
One of the most intriguing implications is the possibility of the decay of our electroweak vacuum. 
Within the Standard Model of particle physics (SM) and with its best-fit parameters, the loop-corrected potential of the SM Higgs field develops large negative values at large field values, making the low-energy electroweak vacuum metastable~\cite{Flores:1982rv,Sher:1988mj,Isidori:2001bm,
1205.2893,1205.6497,1307.3536,Bednyakov:2015sca}.
The fact that the decay into the lower, negative-energy state did not happen in the history of our universe can constrain the parameters of the SM and its extensions; see \cite{Markkanen:2018pdo} for a review.

While in the present-day low-energy, low-temperature background the SM Higgs vacuum is sufficiently long-lived, the situation could be different at other epochs or near extreme objects such as black holes (BHs). The BH catalysis of vacuum decay \cite{Hiscock:1987hn,Berezin:1987ea,Arnold:1989cq,Berezin:1990qs} is an interesting open problem that has received significant interest recently \cite{Gregory:2013hja,Burda:2015isa,Burda:2015yfa,Burda:2016mou,Tetradis:2016vqb,Canko:2017ebb,Gorbunov:2017fhq,Mukaida:2017bgd,Kohri:2017ybt,Dai:2019eei,Hayashi:2020ocn,Miyachi:2021bwd,Briaud:2022few,Strumia:2022jil,Hamaide:2023ayu,He:2024wvt,Rossi:2025fix}. The reason for this is that small evaporating primordial BHs, predicted in many cosmological models \cite{GarciaBellido:1996qt,Fujita:2014hha,Dong:2015yjs,Allahverdi:2017sks,Lennon:2017tqq,Morrison:2018xla,Hooper:2019gtx,Carr:2020gox,Hooper:2020evu}, may drastically enhance the false vacuum decay rate in their vicinity, in particular, due to energetic field fluctuations sourced by the BHs~\cite{Mukaida:2017bgd,Kohri:2017ybt}.
The enhancement occurs at the level of the main exponential suppression part of the rate, and it is important to calculate it accurately.

One of the challenges in calculating the realistic BH induced decay rate is that, in general, a BH is not in thermal equilibrium with the environment (not in the Hartle-Hawking state~\cite{Hartle:1976tp}). In particular, this is true for small, primordial BHs that could form and evaporate in the early universe.
Choosing the correct false vacuum initial state describing such a non-equilibrium system is crucial, since different initial states may have parametrically different lifetimes.
The general non-equilibrium state of interest can be described by the combination of the 
hot source of radiation (BH) with the temperature $\l$ and a comparatively cold background radiation (early-universe plasma) with the temperature $\l'$. In the limit $\l=\l'$ we obtain the BH in equilibrium known as the Hartle-Hawking state \cite{Hartle:1976tp}. In the limit $\l'\to 0$, we obtain the BH radiating in empty space, which is the Unruh state \cite{Unruh:1976db}. 

Semiclassical methods can be applied to calculate the false vacuum decay rate. In the semiclassical limit the decay is described by bounce --- a complex solution of the field equations representing the saddle
point of the Feynman path integral \cite{doi:10.1142/3768}. 
It is well known how to obtain the bounce for systems in equilibrium
\cite{Coleman:1977py,Callan:1977pt,Coleman:1978ae,Coleman:1980aw}. It is a regular Euclidean solution satisfying periodic boundary conditions in the imaginary time. For systems out of equilibrium using Euclidean field configurations for calculating the decay suppression is not justified, and more general methods must be used.
One method is to start 
from the path
integral expression for the transition amplitude in real time from the
false vacuum at $t\to -\infty$ to the true vacuum at $t\to
+\infty$ and deform the contour so that the path integral can be evaluated in the saddle-point approximation, yielding the bounce solution
\cite{Miller,
Rubakov:1992ec,Bonini:1999kj,Bezrukov:2003tg,
Bramberger:2016yog,Hayashi:2021kro}
(see also
\cite{Turok:2013dfa,Cherman:2014sba,Andreassen:2016cff,Andreassen:2016cvx}). 

The in-in formalism has been employed to study the BH catalyzed vacuum decay from the Hartle-Hawking and Unruh states \cite{Shkerin:2021zbf,Shkerin:2021rhy} in models of two-dimensional dilaton gravity and the scalar field with a tunneling potential. It was found that the two vacuum states indeed have parametrically different lifetimes, and that the decay of the Unruh vacuum can be exponentially suppressed at all BH temperatures, contrary to the Hartle-Hawking vacuum. The advantage of using the two-dimensional gravity models is that they allow us to find the tunneling solution and the decay suppression analytically. At the same time, these models can reproduce the essential features of the realistic $4d$ Schwarzschild BH background, such as the near-horizon Rindler region, the far-from-horizon flat region and the centrifugal barrier for the field modes separating the two regions \cite{Shkerin:2021rhy}.

In this work we extend the analysis of Refs.~\cite{Shkerin:2021zbf,Shkerin:2021rhy} to the general non-equilibrium state in which the BH and the thermal environment coexist at different temperatures. It is a natural question how the presence of the environment affects the BH induced decay rate \cite{He:2024wvt}.
It is interesting to study the catalyzing effect of such two-component radiation, both in the BH vicinity and in the asymptotically-flat region.
We will construct the tunneling solution describing the decay from the general state with $\l'\neq \l$ and see how it interpolates between the thermal and Unruh bounces in the respective limits. 
In the limits $\l'=\l$ and $\l'/\l\ll 1$ we recover the Hartle-Hawking and Unruh states, correspondingly. 
In particular, the leading contribution to the Unruh vacuum decay suppression is obtained by taking the limit $\l'/\l\ll1$ in our general result.
The interpolation procedure can be useful to find the Unruh bounce in $4d$, where the analytical treatment is not possible and one must resort to numerical methods. 

One interesting aspect of our analysis concerns with the decays at high temperature(s), where the tunneling bounce solution ceases to exist \cite{Shkerin:2021zbf,Shkerin:2021rhy}, and the decay proceeds via the formation of the sphaleron, which corresponds to the top of the potential barrier between the false and true vacua.\footnote{It has been argued that for a large class of BH solutions, the vacuum decay near a BH in equilibrium always proceeds via the sphaleron \cite{Briaud:2022few}.}  
The transition from the under-barrier tunneling to the tunneling via the sphaleron at a certain energy of the initial state is a general feature of quantum-mechanical systems
\cite{Bezrukov:2003tg,Bezrukov:2003er,Takahashi_2003,
Levkov:2004tf,Levkov:2004ij, 
Takahashi_2005,Levkov:2007yn,Levkov:2008csa,Demidov:2015bua}.
The relevant saddle point of the decay amplitude must be the solution that interpolates between the false vacuum and the sphaleron. 
For systems in equilibrium, the sphaleron is the critical bubble --- a static solution of the Euclidean equations of motion \cite{Linde:1981zj}. 
On the other hand, finding the sphaleron in non-equilibrium systems is a non-trivial problem.
In this work we will construct analytically the sphaleron corresponding to the general non-equilibrium state with $\l'\neq\l$, as well as the interpolating saddle point solution, in a certain range of the parameters of the model. The non-equilibrium sphaleron resembles the thermal sphaleron boosted in the direction of the radiation flux with the higher temperature. The action of such ``flying'' sphaleron gives the correct vacuum decay suppression at high $\l$ and/or $\l'$, 
which can be verified using an estimate based on the stochastic picture. 

The paper is organized as follows. In Sec.~\ref{sec:setup} we describe the model of the scalar field with an unstable potential in the background of the dilaton BH in two dimensions.\footnote{Throughout the paper, we will neglect the back-reaction of the tunneling field onto the BH geometry.}
We introduce the false vacuum state and outline the technique to find the tunneling
solution. 
In Sec.~\ref{sec:tunn} we first illustrate the technique in the case of vacuum decay in flat space. Then we study tunneling in the BH background, both near and far from horizon and at general $\l,\l'$. In Sec.~\ref{sec:sph} we study the decay at high temperature(s) where the relevant solution describes the ``jump'' onto the flying sphaleron. We also estimate the decay rate in this regime based on a simple stochastic picture. 
Combining with the results of Sec.~\ref{sec:tunn}, we then summarize the different decay channels and obtain the whole picture.
In Sec.~\ref{sec:disc} we discuss our findings and conclude. Several appendices contain technical details of the calculations.

\section{Setup}
\label{sec:setup}

\subsection{Theory}
\label{ssec:theory}

We consider a real scalar field with unstable potential, in the background of a dilaton BH in two dimensions \cite{Callan:1992rs}. 
The dilaton BH is characterized by mass $M$ and temperature $T_{\rm BH}\equiv \l/(2\pi)$. 
They are a priori independent of each other; however, to reproduce the behavior of a Schwarzschild BH in four dimensions, in what follows we take the mass to be inversely proportional to the temperature: $M=M_0^2/\l$, where $M_0$ is a constant parameter.

The BH background is set by the metric and the dilaton field $\phi$.
It is convenient to use the conformally-flat ``tortoise'' coordinates $(t,x)$ covering the outer region of the BH,\footnote{As discussed in \cite{Shkerin:2021zbf}, this is the only
region relevant for vacuum decay.} in which the line element takes the form
\be \label{ds2}
\diff s^2 = \O(x)(-\diff t^2+\diff x^2) \;.
\ee
The conformal factor $\O$ and the dilaton field $\phi$ are given by
\be \label{BH_backgr}
\O(x)=\frac{1}{1+\e^{-2\l x}} \;, ~~~ \phi(x) = -\frac{1}{2}\log\left[ \frac{M_0^2}{2\l^2}\left( 1+\e^{2\l x} \right) \right] \;.
\ee
The BH horizon is located at $x\to -\infty$ where $\O(x)\approx \e^{2\l x}$ and the spacetime is described approximately by the Rindler metric. In the opposite limit $x\to\infty$ the spacetime is approximately flat.
See Appendix~\ref{app:dilaton} for more details of the dilaton BH.

The action of the scalar field in the background (\ref{ds2}) is given by
\be \label{S}
S = \frac{1}{\gc^2 } \int \diff t \diff x \left(-\frac{1}{2} (\d\vf)^2 - \O(x) V(\vf) - \frac{1}{2}\l q \O'(x) \vf^2 \right) \;.
\ee
Here $\gc$ is the small coupling ensuring the validity of the semiclassical approximation,\footnote{To avoid back-reaction of tunneling on the BH background, the coupling $\gc$ must be larger than the gravitational coupling $\e^{2\phi}$ \cite{Callan:1992rs} at all $x$. Using \cref{BH_backgr} this leads to the condition $M_0^2\gg \l^2/\gc^2$, which can always be satisfied for large enough $M_0$. } 
the tunneling potential $V(\vf)$ is discussed below, and the third term arises from the nonminimal coupling of $\vf$ to the dilaton, $\mathcal{L}_{\phi\vf}=-Q \e^{2\phi}\vf^2$, upon using \cref{BH_backgr} and identifying $q\equiv2Q/M_0^2$.
This term gives rise to the temperature- (and $x$-)dependent
barrier in the potential for linearized field perturbations, which is similar to the centrifugal
barrier in the $4d$ Schwarzschild spacetime.
This ``dilaton'' barrier has the crucial effect on the Unruh vacuum decay rate \cite{Shkerin:2021rhy}.
From the analysis of the Hartle-Hawking vacuum decay in the theory (\ref{S}) we know that the essential features of the thermal bounce (sphaleron) around Schwarzschild BH are reproduced if the strength of the dilaton barrier $q$ is in the region 
$m^2/(2\l^2)\lesssim q\ll 1$. We expect this to be true for the more general non-equilibrium vacuum studied here. In the next section we will discuss the bound on $q$ in more detail.

\begin{figure}[t]
    \centering
    \includegraphics[width=0.5\linewidth]{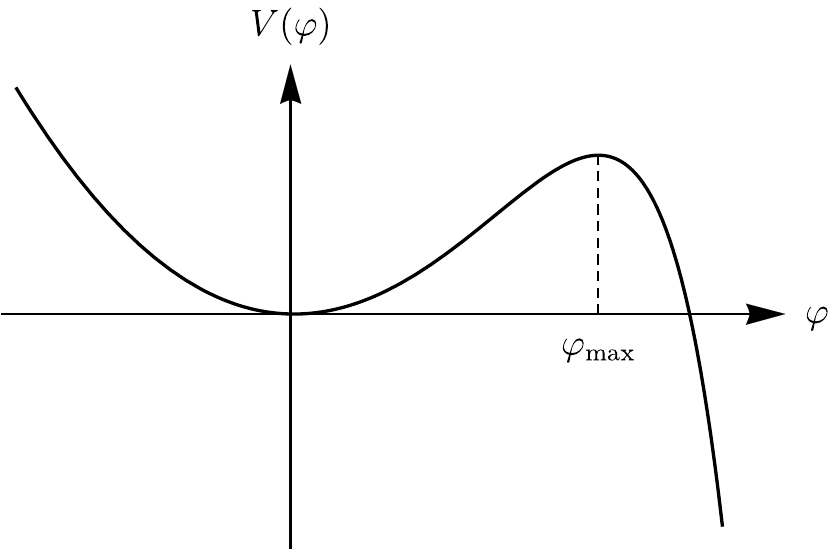}
    \caption{The scalar field potential.}
    \label{fig:pot}
\end{figure}

We take the following scalar field potential:
\be \label{V}
V(\vf) = \frac{1}{2}m^2\vf^2 - 2 \k (\e^\vf-1) \;.
\ee
The potential is shown in Fig.~\ref{fig:pot}.
The false vacuum is at $\vf\approx 0$. The true vacuum is approached at $\vf\to\infty$. The special form of the nonlinear part of the potential $V_{\rm int}(\vf) \equiv - 2\k (\e^\vf-1)$ ensures that the tunneling solution and the corresponding decay suppression can be found analytically, both near the BH ($\Omega(x) \approx \e^{2 \lambda x}$) and far from it ($\Omega(x) \approx 1$), if the following condition is satisfied:
\be \label{hierarchy}
\ln\frac{m}{\sqrt{\k}} \gg 1 \;.
\ee
The maximum of the potential is located at 
\be \label{Phimax}
\vf_{\rm max}\approx \ln\frac{m^2}{2\k} \;.
\ee
Under condition (\ref{hierarchy}), the potential is nearly quadratic at $\vf<\vf_{\rm max}$, while at $\vf>\vf_{\rm max}$ it quickly becomes negative and is dominated by the nonlinear term. Thus, the potential has two energy scales: the mass scale $m$ and the barrier scale $m\ln(m/\sqrt{\k})\gg m$. This hierarchy ensures that the tunneling bounce, which is essential to evaluate the vacuum decay rate, as we will see,  has the narrow nonlinear ``core'' where it probes the region outside the barrier and where the mass term in (\ref{V}) can be neglected, and the broad linear ``tail'' where it approaches the false vacuum and where the nonlinear term can be neglected.
Crucially, in both regions the bounce equation can be solved analytically,\footnote{More precisely, it can be solved analytically in the asymptotically-flat region and the near-horizon region described by the Rindler metric.} as discussed later, and the full solution 
is obtained
by matching its core and tail parts in the overlapped region. 

Finding the bounce solution requires the knowledge of the boundary conditions imposed by the initial false vacuum state. We will presently discuss these conditions.

\subsection{False vacuum state}
\label{ssec:instate}

\begin{figure}[t]
    \centering
    \includegraphics[width=0.5\linewidth]{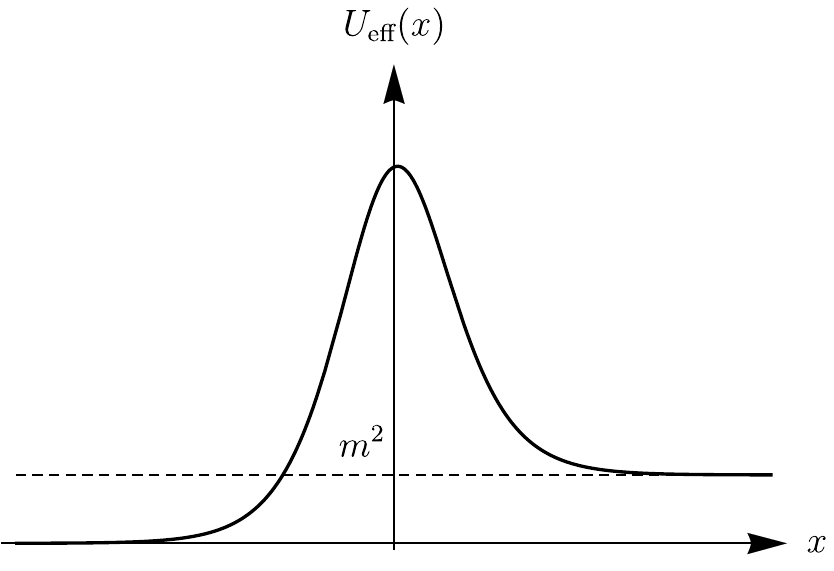}
    \caption{Effective potential for linear field modes in the background of the dilaton BH and with the dilaton barrier $q\gtrsim m^2/(2\l^2)$.}
    \label{fig:Ueff}
\end{figure}

The initial state is specified by providing the occupation numbers of linearized field perturbations around the ground state $\vf\approx 0$. Consider the basis of perturbations made of 
positive- and negative-frequency modes of the field:
\be \label{waves}
\vf^{+}_\o(t,x)=f_\o(x)e^{- i\o
  t}~,~~~~~\vf^{-}_\o(t,x)=f^*_\o(x)e^{i\o t}\; , ~~~ \o>0 \;,
\ee
where the mode functions $f_{\o}(x)$ satisfy the eigenvalue equation
\be\label{EqModes}
-\frac{d^2f_\o}{dx^2}+U_{\rm eff}(x) f_\o=\o^2f_\o \;, ~~ U_{\rm eff}(x) = m^2\O(x) + q\l\O'(x) \;,
\ee
and $\O(x)$ is given in \cref{BH_backgr}.
The effective potential $U_\mathrm{eff}$ is shown in Fig.~\ref{fig:Ueff}. It vanishes at $x \to - \infty$ and approaches to $m^2$ at $x \to \infty$
with a barrier around $x\approx 0$. 
At $\o>m$, \cref{EqModes} has two linearly-independent solutions. It is convenient to choose them describing the left-moving ($f_{L,\o}$) and the right-moving ($f_{R,\o}$) plane waves in the near-horizon region ($x\to-\infty$) and the flat region ($x\to\infty$), correspondingly,
\begin{align}
& f_{L,\o}\propto \e^{-i\o x} \;, ~~ x\to-\infty \;, \label{fL} \\
& f_{R,\o}\propto \e^{ikx} \;, ~~ k=\sqrt{\o^2-m^2} \;, ~~ x\to\infty \;. \label{fR}
\end{align}
The solution $f_{L,\o}$ describes the radiation falling into the BH, while $f_{R,\o}$ describes the radiation directed outwards the BH.
In the presence of the potential $U_{\rm eff}(x)$, the mode function
$f_{L,\o}$ ($f_{R,\o}$) becomes the combination of plane waves at $x\to\infty$ ($x\to-\infty$).
At $\o<m$, there is only one solution to \cref{EqModes}, which is a sum of two plane waves in the near-horizon region and falls off exponentially at $x>0$. We keep the notation $f_{R,\o}$ for this mode, as it is obtained from (\ref{fR}) by the analytic continuation $k\mapsto i\vk$, $\vk\equiv\sqrt{m^2-\o^2}$. We can also formally continue the left-moving mode by setting $f_{L,\o}\equiv 0$ at $\o<m$.

The quantum field is decomposed in the basis of the solutions (\ref{waves}) as follows,
\be
\label{phihat}
\hat\vf(t,x) ={\rm g}\int_0^\infty\!\!\frac{d\o}{\sqrt{4\pi\o}}
\sum_{I=R,L}\big[\hat a_{I,\o}\vf^+_{I,\o}(t,x)
+\hat a_{I,\o}^\dagger \vf^-_{I,\o}(t,x)\big]\;.
\ee 
Here $\hat a$, $\hat a^\dagger$ are the annihilation and creation
operators satisfying the usual commutation relations
\be
\label{acomm}
[\hat a_{R,\o},\hat a_{R,\o'}^\dagger]
=[\hat a_{L,\o},\hat a_{L,\o'}^\dagger]=\delta(\o-\o')\;,
\ee
with all other commutators vanishing. The quantum vacuum state is defined by specifying occupation numbers for the operators $a_R$, $a_L$. We take them as follows,
\be
\label{state}
\langle \hat a_{R,\o}^\dagger \hat a_{R,\o'}\rangle_{\l,\l'} = \frac{\delta(\o-\o')}{\e^{2\pi\o/\l}-1} \;, ~~~~
\langle \hat a_{L,\o}^\dagger \hat a_{L,\o'}\rangle_{\l,\l'}
=\frac{\delta(\o-\o')}{\e^{2\pi\o/\l'}-1} \;.
\ee
The first expression means that the BH radiation is in the thermal state, with the BH temperature $\l$. The second expression means that there is an incoming radiation as well and with the different temperature $\l'$. Thus, \cref{state} describes the general non-equilibrium state of a radiating BH immersed in the external heat bath. 

The vacuum state (\ref{state}) has two notable limits. First, when $\l'=\l$, the BH is in equilibrium with the environment, which corresponds to the Hartle--Hawking vacuum \cite{Hartle:1976tp}. Second, when $\l'\to 0$, the BH radiates in empty space, which corresponds to the Unruh vacuum \cite{Unruh:1976db}. This is the most physically interesting case, as the Unruh vacuum describes a BH formed from a collapse of matter. 
Changing the temperature of the environment from $\l'=\l$ to $\l'=0$ allows us to interpolate between the Hartle--Hawking and Unruh vacua. 
A non-zero $\l' \neq \l$ corresponds to the cosmologically relevant situation in which a BH formed by gravitational collapse coexists with a thermal background whose temperature differs from its Hawking temperature, as expected, {\it e.g.}, in the radiation-dominated Universe.
It is therefore important to understand how the tunneling solution and decay suppression behave in this case. Apart from being an interesting physical problem on its own, such interpolation can give us hints about how to construct the Unruh bounce in four dimensions, where no analytic solution is possible.

\subsection{Bounce and decay rate}
\label{ssec:rate}

\begin{figure}[t]
    \centering
    \includegraphics[width=0.45\linewidth]{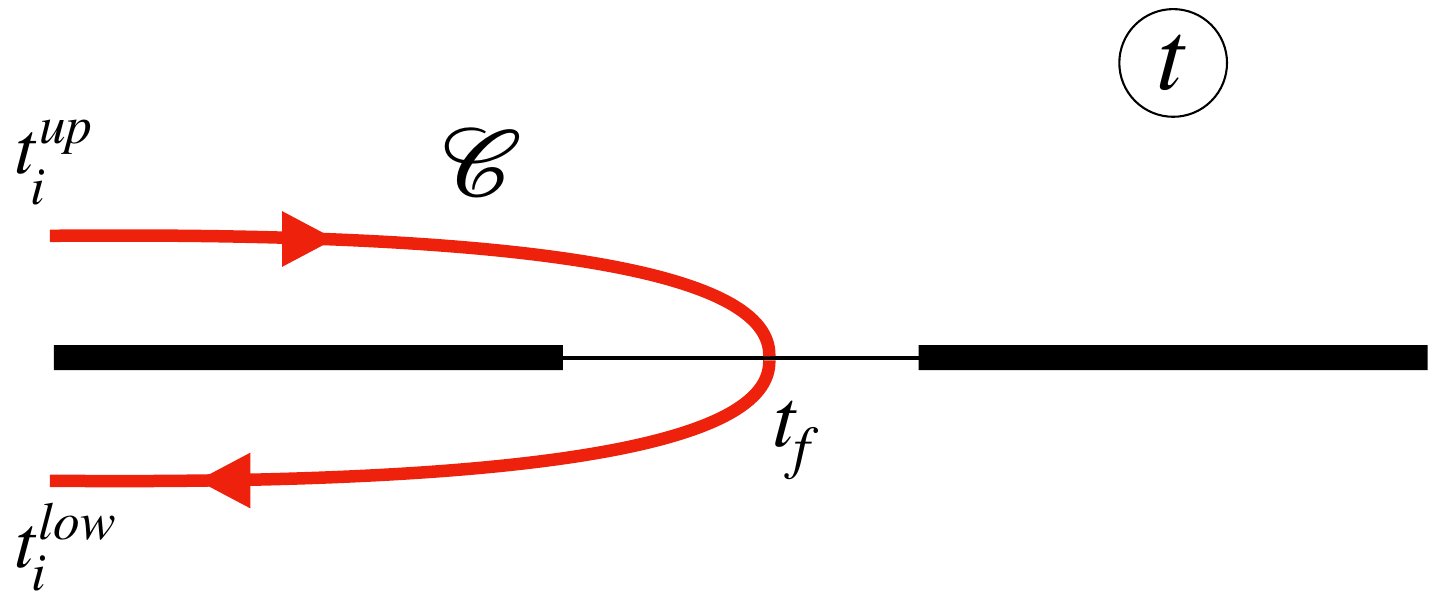}
    \caption{Contour $\C$ in the complex time plane for the
      calculation of the vacuum decay probability. We show the case
      when the branch-cuts of the bounce (shown with thick black
      lines) lie on the real axis. This corresponds to a theory with
      the scalar potential unbounded from below.}
    \label{fig:contour}
\end{figure}

The decay probability is evaluated from the quantum amplitude of transition between the initial false vacuum state introduced in the above and the final state that evolves towards the true vacuum.
We follow the method developed in~\cite{Shkerin:2021zbf}, which is based on the path integral representation of the amplitude in the in-in formalism, and summarize it below.  

In the semiclassical approximation, the decay rate is determined by bounce --- the complex solution of the field equation of motion following from the action (\ref{S}), 
\be \label{EoM}
\left(\Box - U_{\rm eff}(x)\right)\bb - \O (x) V_{\rm int}'(\bb) = 0 \;,
\ee
where $U_{\rm eff}(x)$ is given in \cref{EqModes} and $V_{\rm int}(\vf)$ is the nonlinear part of the potential (\ref{V}).
The bounce must satisfy the 
boundary conditions imposed by the false vacuum state (\ref{state}). To discuss them, we introduce the Green's function corresponding to \cref{EoM}:
\be \label{EoMGreen}
    \left(\Box-U_{\rm eff}(x)\right)\G(t,x;t',x')=i\delta(t-t')\delta(x-x') \;.
\ee
The Green's function can be used to recast \cref{EoM} into the integral form. Selecting a particular solution of \cref{EoM} is equivalent to specifying the Green's function and the contour of integration in the integral equation. 

For the problem at hand, the time integration contour $\C$ is then the one shown in Fig.~\ref{fig:contour}.
It runs from the initial moment of time in the asymptotic past,
$t=t_i^{up}$ shifted to the upper half plane, to the final moment
$t=t_f$ and back to the asymptotic past, $t=t_i^{low}$ in the lower
half plane. The contour
must bypass the singularities of the bounce. Assuming that the bounce is
unique, its values on the upper and lower sides of the contour are
complex conjugated, hence at $t=t_f$ it is real and can be
analytically continued along the real time axis where it describes the
evolution of the field after tunneling. 
We will see below that $\C$ is suitable to describe tunneling in the BH background, but not suitable to describe decays at large $\l$ and/or $\l'$, which occur via classical stochastic jumps of the field over the barrier. To apply the semiclassical method to the high-temperature decays, the integration contour must be modified.

The bounce solution must linearize when $\Re t_i^{\rm up/low}\to-\infty$, and in this limit the bounce is decomposed into the eigenmodes around the false vacuum,
\begin{equation}
    \bb|_{\mathrm{Re}\ t_i^{\rm up}\rightarrow -\infty}=\int_{0}^{\infty}\frac{\mathrm{d}\o}{\sqrt{4\pi\o}}\ \sum_{I=R,L}[c_{I,\o}^{\rm up}\vf^+_{I,\o}(t_i^{\rm up},x)+\bar{c}_{I,\o}^{\rm up}\vf_{I,\o}^{-}(t_i^{\rm up},x)]\ ,
\end{equation}
and, similarly for $\bb|_{\mathrm{Re}\ t_i^{\rm low}\rightarrow-\infty}$ with the coefficients $c_{I,\o}^{\rm low}$, $\bar{c}_{I,\o}^{\rm low}$.
The state (\ref{state}) imposes the following relations between the coefficients,
\begin{equation}\label{boundarycondition0}
    c_{I,\o}^{\rm up}=r_{I}(\o)c_{I,\o}^{\rm low}, \quad r_I(\o)\bar{c}_{I,\o}^{\rm up}=\bar{c}_{I,\o}^{\rm low}\ ,
\end{equation}
where
\begin{equation}\label{boundarycondition}
    r_R(\o)=\e^{-2\pi\o/\l}, \quad r_L(\o)=\e^{-2\pi\o/\l'}\ .
\end{equation}
In the limits $\l=\l'$ and $\l'\rightarrow0$, we recover the boundary conditions for the Hartle-Hawking and Unruh vacua, respectively.
The Green's function satisfying the conditions (\ref{boundarycondition0}), (\ref{boundarycondition}) is an expectation average of the 
linear field operators $\hat{\vf}$ in the state (\ref{state}), ordered along the contour $\C$ \cite{Shkerin:2021zbf}:
\be\label{GreenGen} 
    \G_{\l,\l'}(t,x;t',x') \equiv \frac{1}{\gc^2}\langle
    T[\hat{\vf}(t,x)\hat{\vf}(t',x')]\rangle \;. 
\ee
Its explicit form is derived in Appendix.~\ref{app:Green}.
Overall, the bounce solution $\vf_{\rm b}$ is 
determined by the following integral equation,
\be\label{IntEq}
    \vf_{\rm b} (t,x) = - \mathrm{i} \int_{\C} \diff t' \int_{-\infty}^\infty \diff x' \G_{\l,\l'} (t,x;t',x') \O(x') V_\mathrm{int}'(\vf_\mathrm{b}(t',x')) \;,
\ee 
and at the turning point $t=t_f$ of $\C$ the solution represents the field configuration after the tunneling.

The tunneling rate $\Gamma$ is
defined as the probability of tunneling per unit time.\footnote{Since the background (\ref{BH_backgr}) is $x$-dependent, there is no zero mode associated with the space translations.} We are
interested in the main exponential suppression of decays and write 
\begin{equation}
\label{DecayRate}
    \Gamma\sim \e^{-B} \;.
\end{equation}
In the semiclassical limit,
the coefficient $B$ 
obtained from the path integral for the transition amplitude 
turns out to be
the sum of the imaginary part of the bounce action
computed along the contour $\C$ and a boundary term representing the
initial-state contribution. One can show that the boundary term
cancels upon integration by parts in the action and one is left with
\cite{Shkerin:2021zbf} 
\begin{equation}
\label{B_gen}
    B=-\frac{i}{\gc^2}\int_{\C}\diff t\int_{-\infty}^{\infty}\diff x \left( \frac{1}{2}\bb V_{\rm int}'(\bb)-V_{\rm int}(\bb) \right) \;.
\end{equation}
We see that only
the region where the bounce solution is nonlinear contributes to the suppression.
While the bounce $\bb$ and the suppression $B$ cannot, in general, be found analytically,
the analytic solution can be constructed approximately in the asymptotically-flat and near-horizon (Rindler) regions of the BH using the split-and-match procedure as explained above.

\section{Decay via quantum tunneling}
\label{sec:tunn}

\subsection{Tunneling from Minkowski vacuum}
\label{ssec:Mink}

We first construct the bounce describing tunneling from the flat space at zero temperature, in order to compare it with the tunneling in the BH background.
For convenience, we deform the integration contour $\C$ shown in Fig.~\ref{fig:contour} by rotating its two parts to the imaginary time domain. This is justified as long as we do not cross singularities of the bounce solution. 
The core satisfies equation (\ref{EoM}) with $\O\equiv 1$ and mass term being neglected:
\be\label{EoMCore}
\Box\left.\vf\right\vert_{\rm core} +2\k\e^{\left.\vf\right\vert_{\rm core}} =0 \;.
\ee
On the other hand, the tail of the bounce satisfies the linearized equation of motion, and it must be proportional to the Green's function (\ref{GreenGen}), which in flat space and $\l=\l'=0$ becomes the Feynman Green's function,
\be\label{Green_Feynman}
\G_{F}(-i\tau,x;0,0)=\frac{1}{2\pi}K_0\left( m\sqrt{x^2+\tau^2+i\epsilon} \right) \;,
\ee
where $K_n(x)$ is the modified Bessel function of the second kind and $\tau=it$ is the Euclidean time coordinate.
That this is the case can be seen from \cref{IntEq}: away from the core, the nonlinear term 
is non-vanishing only around the core, and hence
can be approximated by the delta-function, $V_{\rm int}'(\vf_{\rm b}(t',x'))\propto \delta^{(2)}(t',x')$.\footnote{Without loss of generality, we have chosen the center of the bounce be located at $(t,x)=(0,0)$.}
Thus, $\vf_{\rm b}(t,x)$ becomes proportional to the corresponding Green's function.

The general solution of \cref{EoMCore} (Liouville equation) is 
\be \label{CoreGen}
\vf\vert_{\rm core} = \ln \left[ \frac{4F'(-u)G'(v)}{(1+\k F(-u)G(v))^2} \right] \;,
\ee
where $u=t-x$, $v=t+x$, $F(-u)$, $G(v)$ are arbitrary functions and and primes stand for the derivatives of these functions with respect to their arguments.
To match the core with the Green's function (\ref{Green_Feynman}), one should choose\footnote{Intuitively, $F(-u)$ can be associated with the right-moving modes of the linearized core, while $G(v)$ is associated with the left-moving modes. This intuition is useful when considering the general non-equilibrium initial state.}
\be \label{FG_flat}
F(-u)=-C_M u \;, ~~~ G(v) = C_M v \;,
\ee
where $C_M$ is a constant determined by matching procedure.
Introducing the Euclidean radial coordinate $\rho=\sqrt{\tau^2+x^2}$, the core (\ref{CoreGen}) takes the form
\be \label{CoreRho}
\bb\vert_{\rm core} =
\ln\left[\frac{4C^2_{M}}{(1+\kappa C_M^2\rho^2)^2}\right] \;.
\ee
Next, we expand the core at $\rho\gg (C_M\sqrt{\k})^{-1}$ and the Green's function (\ref{Green_Feynman}) at $\rho\ll m^{-1}$. Matching the two expansions at $ (C_M\sqrt{\k})^{-1} \ll \rho \ll m^{-1}$, we find $C_M=m^2/(2\k)\e^{2\g_E}$
where $\g_E$ is the Euler constant. The tail is now determined as $\vf|_\mathrm{tail}=8\pi \G_{F}(-i\tau,x;0,0)$. Crucially, the hierarchy (\ref{hierarchy}) ensures that $ (C_M\sqrt{\k})^{-1} \ll m^{-1}$ and both expansions are valid in some region of $\rho$. Note that the bounce we found is spherically symmetric in the Euclidean spacetime, as it should \cite{Coleman:1977th,Blum:2016ipp,0806.0299}.
Also, the only singularities of the bounce are located on the real-time axis, 
$t_s(x)=\pm (x^2+(\kappa C_M^2)^{-1})^{1/2}$ (see Fig.~\ref{fig:contour}), which justifies the rotation of the contour into the imaginary time domain.

The tunneling suppression is given by \cref{B_gen} where one should substitute the core of the bounce (\ref{CoreRho}) for the dominant contribution~\cite{Shkerin:2021zbf}. We obtain\footnote{The corrections to \cref{B_M} are of order $\gc^{-2}\times o(1)$.}
\begin{equation}
\label{B_M}
    B_M=\frac{16\pi}{\gc^2}\left( \ln\frac{m}{\sqrt{\k}}+\g_E-1 \right) \;.
\end{equation}
We observe that the suppression is enhanced by the large logarithm (\ref{hierarchy}).
Note that the flat-space bounce and suppression (\ref{B_M}) can be calculated using the standard approach, where the bounce is the saddle point of the Euclidean partition function which satisfies the vanishing boundary conditions at $\tau\to\pm\infty$ \cite{Coleman:1977py,Callan:1977pt,Coleman:1978ae}.

\subsection{Tunneling far from horizon}
\label{ssec:tunnfar}

We begin investigating the tunneling from the non-equilibrium initial state (\ref{state}). 
Let us consider first the case where the core is located
far from the BH horizon where the spacetime is approximately flat. 
We will consider the opposite case in the next subsection.
The state (\ref{state}) represents the superposition of two fluxes of thermal radiation traveling in the opposite directions. Studying the catalyzing effect of this two-component radiation is an interesting problem on its own. In our case,
one can think of the left-moving flux as emitted by a black body with temperature $\l'$ located at asymptotic infinity. 
On the other hand, the right-moving flux is the BH Hawking radiation with temperature $\l$, which originates in the near-horizon region and which is moderated by the mode potential $U_{\rm eff}(x)$, see \cref{EqModes}. 
The mode potential also leads to the mixing between the two fluxes, as can be seen by inspecting the general form of the Green's function $\G_{\l,\l'}$ given in Appendix~\ref{app:Green}. Thus, even far from the BH the bounce solution (namely, its linear tail) and the decay suppression are sensitive to the BH geometry and the strength of the dilaton barrier.\footnote{This is the feature of the $(1+1)$-dimensional spacetime where the flux does not dissipate with the distance.}

To find the bounce, we follow the strategy described in Sec.~\ref{ssec:Mink}. 
In the asymptotically-flat region, the equation for the core is the same as that of eq.~\eqref{EoMCore}.
The tail, on the other hand, is proportional to the Green's function $\G_{\l,\l'}$.
The core solution is given by eq.~\eqref{CoreGen} specified by the functions $F(-u)$, $G(v)$  which should be chosen so that the expansions of the core and of the Green's function $\G_{\l,\l'}$ match in the overlapping region. We make the following educated guess:
\be\label{FG_far}
    F(-u)=\frac{C}{\lambda}(\mathrm{e}^{-\lambda u}-d_1) \;,\quad \quad G(v)=\frac{C}{\lambda'}(\mathrm{e}^{\lambda'v}-d_2)\;,
\ee
where $C$, $d_1$ and $d_2$ are constants. We substitute this to the general solution (\ref{CoreGen}) and after some manipulations 
find the following expression,
\be\label{Core_far}
    \bb|_{\text{core}}=\ln\left\{ \frac{\l\l'b_{\rm far}}{\k\{ \ch[ \bl (x-x_c) - \dl t ]-\sqrt{1-b_{\rm far}}\ch[ \bl t - \dl(x-x_c) ]  \}^2} \right\}\ \;,
\ee
where we have introduced the notations 
\be\label{bldl}
    \bl\equiv\frac{\l+\l'}{2},\quad \dl\equiv\frac{\l-\l'}{2}\ , 
\ee
and $x_c$ and $b_{\rm far}$ are determined by $d_1$, $d_2$ and $C$.
Further, $x_c$ is the position of the core at $t=0$, 
and we used the time translation invariance to set $t_f=0$.
The parameter $x_c$ can be chosen arbitrarily as soon as it is in the flat region, $x_c>0$. We further require the matching to the Green's function be performed in the flat region as well, $x_cm\gg 1$.
The parameter $b_{\rm far}$ is determined from matching to the Green's function; we will see that one needs $b_{\rm far}< 1$ for the successful matching.

We make several observations regarding the form of the bounce core (\ref{Core_far}):

\textit{(i)} In general, the bounce is not periodic in the imaginary time. This is expected, since it describes tunneling from the out-of-equilibrium state, and the standard boundary condition of periodicity in the imaginary time is not applicable.

\textit{(ii)} For real $(x,t)$, the bounce is real for $b_\mathrm{far} \leqslant 1$. Its singularities are located at 
\be \label{Bounce_sings}
\hat{t} =\hat{t}_s(\hat{x})=   -\frac{1}{\bl}\left\{\dl x_c \pm \mathop{\rm arcch}\nolimits\left[ \frac{\ch [ \bl (\hat{x}-x_c) ]}{\sqrt{1-b_{\mathrm{far}}}} \right]\right\} \;,
\ee
where we introduced the rotated coordinates
\be \label{CoordsRot}
\bl\hat{x}=\bl x - \dl t \;, ~~~~ -\bl\hat{t}=\dl x - \bl t \;.
\ee
The singularities prevent us from going arbitrarily to the imaginary time domain, contrary to the previous Minkowski case, and we must work with the original contour $\C$.

\textit{(iii)} In the limit when the two fluxes have equal temperature, $\l=\l'$, the solution (\ref{Core_far}) becomes
\be \label{Core_far_th}
    \bb|_{\text{core}}=\ln\left\{ \frac{\l^2 b_{\rm far}}{\k\{\ch[ \l (x-x_c) ]-\sqrt{1-b_{\rm far}}\ch [\l t] \}^2} \right\}\ \,.
\ee
It is now periodic in the imaginary time with the period $2\pi/\l\equiv 1/T_{\rm BH}$. Thus, we recover the thermal bounce describing the decay of the Hartle-Hawking vacuum far from the BH. 
As discussed in \cite{Shkerin:2021zbf,Shkerin:2021rhy}, we can deform the contour $\C$ to contain the imaginary time segment $-\pi/\l\leqslant \tau\leqslant\pi/\l$ and apply the standard Euclidean prescription to compute the bounce action. 

\textit{(iv)} The limit $\l'\to0$ is more tricky. In this limit we expect to recover the flat Unruh bounce describing tunneling from the Unruh vacuum far from the BH, which was studied in \cite{Shkerin:2021zbf}.\footnote{Ref.~\cite{Shkerin:2021zbf} studies vacuum decay in the theory without the dilaton barrier, $q=0$. However, this does not affect the form of the core, only its matching with the Green's function.} 
We cannot take this limit directly in \cref{Core_far}, the reason being that we neglect the mass term when deriving $\bb|_{\rm core}$, which implicitly assumes $\l,\l'\gg m$. 
To proceed, we redefine $d_2\equiv \e^{\l'v_1}$ in \cref{FG_far}, where $v_1$ is a new constant. Then $G(v)\to C (v-v_1)$ at $\l'\to0$, meaning that the left-moving modes are now in vacuum, cf. \cref{FG_flat}. 
The solution takes the form 
\be\label{Core_far_U1}
\begin{split}
   & \bb|_{\text{core}}=\\
   & \ln\left\{ \frac{\lambda\lambda'b_{\rm far}}{\k\{\ch[\frac{\lambda'}{2}(v-v_c)-\frac{\lambda}{2}(u-u_c)+\frac{1}{2}\ln(1-b_{\rm far})]-\sqrt{1-b_{\rm far}}\ch[\frac{\lambda'}{2}(v-v_c)+\frac{\lambda}{2}(u-u_c)]\}^2} \right\}\ ,
\end{split}
\ee
where $d_1$, $v_1$ and $C$ are absorbed in $u_c$, $v_c$ and $b_{\rm far}$. 
Next, we replace $\l'\mapsto m$ and expand in the small parameters $m|v-v_c|$, $m/\l\ll 1$. To the leading order the result is
\be\label{Core_far_U2}
    \bb|_{\text{core}}=\ln\left\{\frac{4\l^2b_U}{\kappa\{-2\l(v-v_c)\sh[\frac{\lambda}{2}(u-u_c))]+b_U\e^{\frac{\l}{2}(u-u_c)}\}^2}  \right\}\ ,
\ee
where we denoted $b_U=\l b_{\rm far}/m$.
This matches the form of the Unruh bounce found in \cite{Shkerin:2021zbf}.

\begin{figure}[t]
    \centering
    \includegraphics[width=0.99\linewidth]{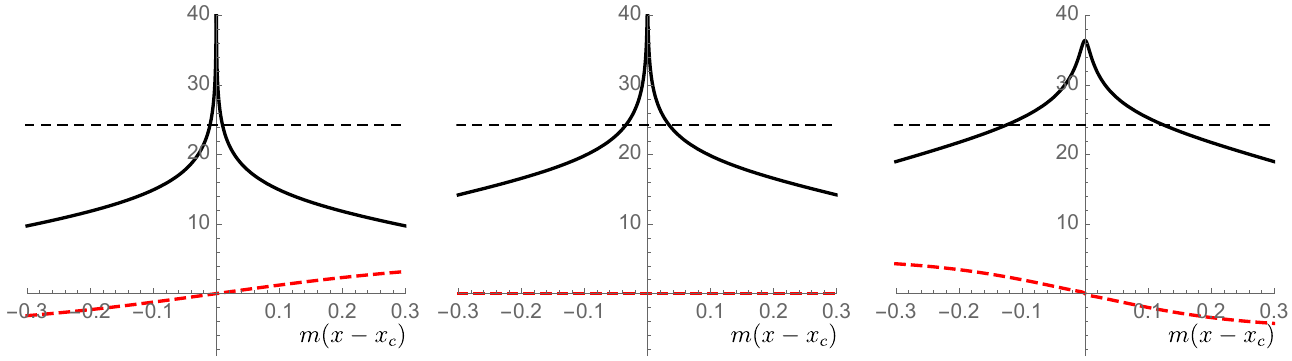}
    \caption{Bounce solution (\ref{Core_far}) describing tunneling from the non-equilibrium vacuum (\ref{state}) far away from the BH.
    Solid black line is the bounce profile $\bb$, the red dashed line is its time-derivative $\dot{\bb}/m$, both are shown at the nucleation time $t=0$. The gray dashed line marks the field value $\vf_{\rm max}$
at the maximum of the potential barrier, see \cref{Phimax}.
We take $\ln(m/\sqrt{\k})=25$, the BH temperature $\l/m=10$, the dilaton barrier $q\l/m=1$ and the temperature of the environment $\l'/m=5$ \textbf{(left)}, $\l'/m=10$ \textbf{(middle)}, $\l'/m=15$ \textbf{(right)}.
}
    \label{fig:bounce_far}
\end{figure}

\textit{(v)} Comparing the general non-equilibrium solution (\ref{Core_far}) with the thermal solution (\ref{Core_far_th}), we see that (\ref{Core_far}) describes the bounce core moving with the velocity $V_c=\dl/\bl$ in the same direction as the flux with the higher temperature. The coordinates (\ref{CoordsRot}) are the rest frame of the core. 
Thus, we can think of the non-equilibrium bounce core as the boosted thermal bounce core. This intuition is particularly useful when considering sphaleron solutions describing high-temperature decays, see Sec.~\ref{sec:sph}. Note, however, that the full bounce solutions, including the linear tails, are not related by a simple Lorentz boost, since the mode potential introduces the mixing between the two fluxes.
Figure~\ref{fig:bounce_far} illustrates the bounce profile and its time-derivative at the moment of nucleation for different values of the ratio $\l/\l'$. We observe the motion of the nucleation center in the direction of the radiation with the higher temperature.

\begin{figure}[t]
    \centering
    \includegraphics[width=0.35\linewidth]{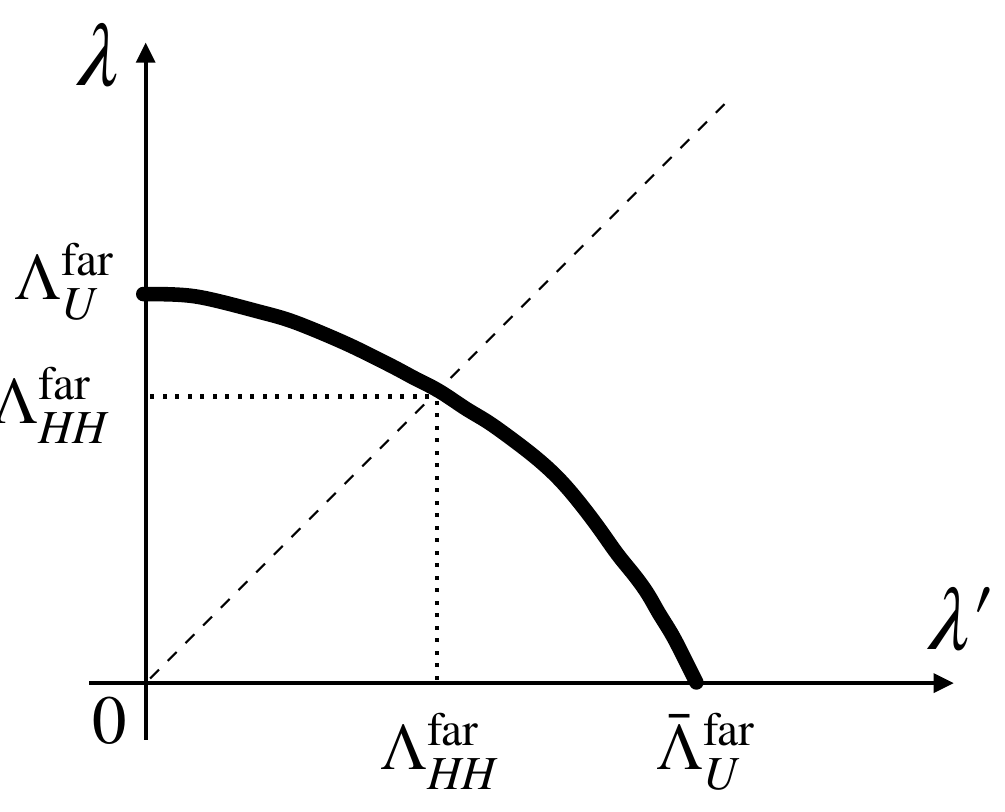}
    \caption{Critical line $b_{\rm far}(\l,\l')=1$. The dashed straight line is $\l=\l'$.}
    \label{fig:crit_far}
\end{figure}

We now match the core (\ref{Core_far}) to the linear tail of the bounce. Assume that $b_{\rm far}\ll 1$ and expand (\ref{Core_far}) in the region
\be 
\ch[ \bl (x-x_c) - \dl t ] \gg b_{\rm far} \ch[ \bl t - \dl(x-x_c) ] \;, \label{far_hr_core_exp_region}
\ee
while still assuming the close separation of points, 
$m|x-x_c|$, $m|t|\ll 1$.\footnote{The properties of the matching region along the contour $\C$ are discussed in more detail in \cite{Shkerin:2021zbf}.}
In this region we match the core expansion 
to the Green's function in
eq.~\eqref{rightgreenfunction1}. 
The matching gives
\be \label{Tail_far}
\bb|_{\rm tail} = 8\pi\G_{\l,\l'}^{\rm far}(t,x;0,x_c) 
\ee 
and fixes $b_{\rm far}$ as follows,
\be \label{b_far}
b_{\rm far}=\frac{\kappa}{4\lambda\lambda'}\mathrm{e}^{4\frac{\lambda-\lambda'}{m\pi}\tilde{\mathcal{H}}(\frac{q\lambda}{m})+\frac{2\lambda'}{m}} \;,
\ee
where $\tilde{\H}(y)$ is a monotonically decreasing function of $y$, with $\tilde{\H}(0)=2/3$ and $\tilde{\H}(y)\approx \pi/(2y)$ as $y\rightarrow \infty$; the explicit form of this function is given in \eqref{htilde}.  
The parameter $b_{\rm far}$ grows when any of the temperatures $\l$, $\l'$ grows. 
Figure~\ref{fig:crit_far} shows the critical line $b_{\rm far}(\l,\l')=1$.
As discussed in the next section, the matching region can be extended from $b_{\rm far}\ll 1$ up to the critical line. 
Note that the hierarchy between $m$ and $\sqrt{\kappa}$ (eq.~\eqref{hierarchy}) guarantees the existence of the critical line and ensures that the critical temperatures are well above the mass of the field.

The critical line 
marks the boundary of the validity region of the bounce solution (\ref{Core_far}), (\ref{Tail_far}), when the factor $\sqrt{1-b_{\rm far}}$ in eq.~\eqref{Core_far} becomes imaginary. 
The physical interpretation of this phenomenon was discussed in \cite{Shkerin:2021zbf,Shkerin:2021rhy}, albeit for the equilibrium and Unruh states only. At temperatures much below the critical line, $b_{\rm far}(\l,\l')\ll1$, the vacuum decay occurs via quantum tunneling, and the bounce (eq.~(\ref{Core_far}) for the core and  eq.~(\ref{Tail_far}) for the tail) is the relevant semiclassical solution. At and above the critical line, the dominant decay channel is via classical stochastic jumps of the field over the barrier. For the equilibrium state, the relevant semiclassical solution in this case is the sphaleron (or the critical bubble). For the theory at hand, it was constructed in \cite{Shkerin:2021rhy}.\footnote{More precisely, the relevant solution interpolates between the thermal vacuum in the asymptotic past, $\Re t\to-\infty$, and the static sphaleron in the asymptotic future, $\Re t\to\infty$, see Sec.~\ref{sec:sph}.}
For a non-equilibrium state, no such solution has been constructed analytically in a field theory, to the best of our knowledge.\footnote{See \cite{Levkov:2004tf,Demidov:2011dk,Demidov:2015bua,Demidov:2015nea} where similar ``jump-onto-the-sphaleron'' solutions were found numerically to describe creation of solitons in particle collisions and collision-induced tunneling. } 
We will do this in Sec.~\ref{sec:sph}.

In general, the equation $b_{\rm far}(\l,\l')=1$ cannot be solved analytically to find one critical temperature $\L$ as a function of another, $\L'$. It is possible to find explicit expressions for $\L,\L'$ in several cases:
in the equilibrium regime when $\L=\L'\equiv\L_{HH}^{\far}$,
in the Unruh vacuum limit with $\L\equiv \L_U^{\rm far}$, $\L'\to0$, and in the limit of cold BH $\L\to0$, $\L'\equiv\bar{\L}_{U}^{\rm far}$. To the leading-log accuracy, we find
\be\label{TempCritFar}
\L_{HH}^{\far} \simeq m\ln\frac{m}{\sqrt{\k}} \;, ~~~~ \L_{U}^{\far} \simeq \frac{m\tilde{y}_c}{q} \;, ~~~~ \bar{\L}_{U}^{\rm far} \simeq \frac{m}{1-\frac{4}{3\pi}}\ln\frac{m}{\sqrt{\k}} \;,
\ee
where $\tilde{y}_c$ is the solution of the equation $\tilde{y}_c\tilde{\mathcal{H}}(\tilde{y}_c)=a\pi/2$ and
we introduced a new parameter
\be \label{q_a}
a = q \ln \frac{m}{\sqrt{\k}} \;.
\ee
The solution exists at sufficiently small $a$ due to the asymptotics of the function $\tilde{\mathcal{H}}$, see eq.~\eqref{htilde}. In what follows we choose the scalar-dilaton coupling $q$ such that $a<1$.
As discussed in \cite{Shkerin:2021rhy}, for this choice of the coupling the behavior of the bounce solution and the decay suppression in the Hartle-Hawking case is qualitatively the same as for the $4d$ Schwarzschild BH. We expect this to be true also for the general non-equilibrium state. Note that the critical temperatures (\ref{TempCritFar}) are always enhanced by the large logarithm $\ln(m/\sqrt{\k})$.

Turning to the tunneling suppression, we need to calculate the integral (\ref{B_gen}) on complex-time contour $\C$ shown in Fig.~\ref{fig:contour}. Unlike
Minkowski or flat-space thermal cases, we cannot deform the contour to cast this integral into the form
of an Euclidean action, and we have to work with the original contour $\C$.
The calculation is presented in Appendix~\ref{app:action} and the result is
\begin{equation}\label{B_tunn_far}
    B_{\far}^{\rm tunn.}=\frac{16\pi}{\gc^2}\left(\ln{\sqrt{\frac{\lambda'\lambda}{\kappa}}}-\frac{\lambda'}{2m}-\frac{\lambda-\lambda'}{m\pi}\tilde{\mathcal{H}}(\frac{q\lambda}{m})+\ln{2}-1\right) \ .
\end{equation}

Let us examine the consistency and implications 
of this result.
First, it is straightforward to check that at $\l=\l'$ we obtain the flat-space thermal bounce suppression studied in \cite{Shkerin:2021zbf}. Interestingly, the suppression is not sensitive to the mode potential generated by the BH geometry. In particular, it does not depend on the dilaton barrier $q$. This is expected, since the Hartle-Hawking state reduces to the flat-space thermal state far from the horizon.
Next, even though we assume $\l,\l'\gg m$ in calculating the action, one can formally set 
$\l'\mapsto m$ in (\ref{B_tunn_far}) and, disregarding the order-one numbers, obtain the leading-log and the linear in $\l/m$ terms in the Unruh bounce suppression in the asymptotically-flat region. In this case, it is sensitive to the dilaton barrier moderating the flux of Hawking radiation. In particular, since $\tilde{\mathcal{H}}(y)$ is a monotonically decreasing function of $y$, the decay is more suppressed for larger values of $q$. The sensitivity to the BH geometry and the dilaton barrier remains for a general $\l'\neq\l$.
The barrier also affects the incoming radiation flux, reflecting part of it back to the asymptotically-flat region. This leads to the interesting effect: if the BH is colder than the environment, $\l<\l'$, increasing $q$ enhances the decay probability, since more radiation is reflected back.

\subsection{Tunneling near horizon}
\label{ssec:tunn_near}

Next, we consider the case when the core of the bounce is located in the near-horizon region.
In the vicinity of the BH the vacuum decay is affected both by field excitations and nontrivial geometry. The near-horizon region is approximately described by the Rindler metric with the conformal factor $\O=\e^{2\l x}$. As before, we solve the bounce equation separately in the nonlinear regime, where the mass term can be neglected, and the linear regime. Taking into account the conformal factor, the nonlinear core satisfies 
\be\label{EoMCore_near} 
\Box \bb\vert_{\rm core} + 2\k \e^{2\l x + \bb\vert_{\rm core}} = 0 \;.
\ee
The general solution can still be found exactly in terms of two arbitrary functions $F(-u)$, $G(v)$, cf. \cref{CoreGen}:
\be \label{CoreGen_near}
\bb\vert_{\rm core} = \ln \left[ \frac{4F'(-u)G'(v)}{\O(u,v)(1+\k F(-u)G(v))^2} \right] \;. 
\ee
Importantly, the core must fit entirely in the Rindler region of the BH for this solution to be valid. We will see that this is the case for $\l,\l'$ below a certain critical line.
Once more, the linear tail of the bounce is determined by the Green's function $\G_{\l,\l'}$, but computed in the BH vicinity, see Appendix~\ref{app:Green}. Matching with the Green's function determines the functions $F(-u)$, $G(v)$. It turns out that these functions have the same form as for the asymptotically-flat region, \cref{FG_far}. 
Thus, we obtain
\be\label{Core_near}
    \bb|_{\text{core}}=\ln\left\{ \frac{\l\l'b_\mathrm{near} }{\k\{ \ch[ \bl (x-x_c) - \dl t ]-\sqrt{1-b_\mathrm{near}}\ch[ \bl t - \dl(x-x_c) ]  \}^2} \right\}\ - 2 \l x \;,
\ee
where $\bl,\dl$ are defined in (\ref{bldl}). The linear term arises from the nontrivial BH geometry.
The core fits the near-horizon region provided that $\l|x_c|\gg 1$, $x_c<0$.
As in the far-from horizon case, the matching to the tail of the bounce, which is proportional to the Green's function, is easy to perform if $b_\mathrm{near} \ll 1$. As discussed in Sec.~\ref{sec:sph}, the matching can in fact be extended to larger values $b_\mathrm{near} \leqslant 1$.

Let us analyze the solution (\ref{Core_near}), in particular, how it is different from the far-from-horizon bounce. 
The comments \textit{(i), (ii)} to the solution (\ref{Core_far}) apply to (\ref{Core_near}) without change. The comments \textit{(iii)} to \textit{(v)} are modified as follows. 

\textit{(iii$\:'$)} In the case of the BH in equilibrium with the environment, $\l=\l'$, the solution reduces to \cref{Core_far_th} with the linear term $-2\l x$ added, and we recover the near-horizon bounce for the Hartle-Hawking vacuum. 
It has the same periodicity in the imaginary time. Moreover, it is the thermal bounce core in the Rindler spacetime, which can be transformed to the vacuum bounce core (\ref{CoreRho}) by a suitable change of variables $(t,x)$~\cite{Shkerin:2021zbf}. That such transformation is possible is the consequence of the fact that the Rindler metric describes the flat spacetime as seen by uniformly accelerating observers. Note that there is no such correspondence between the linear tails of the two bounces, since the tail probes the spacetime nonlocally. 

\textit{(iv$\:'$)} Taking the limit of small $\l'$, $\l' \rightarrow m$ while $m/\l \ll 1$, in the same way as for the far-from-horizon bounce, one can recover the near-horizon Unruh bounce studied in \cite{Shkerin:2021rhy}.

\textit{(v$\:'$)} As in the far-from-horizon case, the solution (\ref{Core_near}) describes the core moving in the direction of the flux with the higher temperature. We are free to choose the core position $x_c$ at $t=0$, as long as it is deep in the Rindler region, $\l|x_c|\gg 1$ and $x_c<0$. This freedom is again a consequence of the mapping between the Rindler and flat spacetime. The BH corrections to the Rindler metric are expected to lift the degeneracy and pick one, least-action solution.

Let us match (\ref{Core_near}) to the Green's function computed in the Rindler region at close separation. We assume $b_{\rm near}\ll 1$ and expand (\ref{Core_near}) at
\be 
\ch[ \bl (x-x_c) - \dl t ] \gg b_{\rm near} \ch[ \bl t - \dl(x-x_c) ] \;, \label{near_hr_core_exp_region}
\ee
and $m|x-x_c|$, $m|t|\ll 1$. Comparing with \cref{leftgreenfunction1}, we obtain
\be\label{near tail}
    \bb|_{\text{tail}}=8\pi\G_{\l,\l'}^{\rm near}(t,x;0,x_c) \;
\ee
and find the parameter $b_{\rm near}$:
\begin{equation}\label{b_near}
    b_{\near}= \bar{b}_{\near}\e^{-2\l x_c} \;, ~~~ \bar{b}_{\near}= \frac{\k}{4\l\l'}\e^{4\frac{\l-\l'}{m\pi}\H(\frac{q\l}{m})+\frac{4\l'}{m+q\l}} \;, 
\end{equation}
where 
$\H(y)$ is a monotonically decreasing function of $y$, with $\H(0)=8/3$ and $\H(y)\approx \pi/(2y)$ as $y\rightarrow \infty$;
the explicit form of this function is given in \eqref{hq}.
Similarly to the far-from-horizon case, we define the critical line ${\bar b}_{\rm near}(\l,\l')=1$, above which the matching becomes impossible for the near-horizon solution at any $x_c<0$, i.e. the core stops fitting the Rindler region of the BH (or $\sqrt{1-b_{\rm near}}$ cannot be real).\footnote{That this is true for all values of $\l'/\l$, including the Hartle-Hawking and Unruh vacua, is ensured by the smallness of the dilaton barrier coupling $q$. See \cite{Shkerin:2021rhy} for the discussion of the opposite, less physically interesting case for the Hartle-Hawking vacuum.} 
It is straightforward to find the critical temperatures in the Hartle-Hawking, Unruh and ``anti-Unruh'' vacua 
(cf. \cref{TempCritFar}):
\be \label{TempCritNear}
\L_{HH}^{\rm near} \simeq \frac{m\ln\frac{m}{\sqrt{\k}}}{2-a}  \;, ~~~~ \L_U^{\rm near} \simeq \frac{y_cm}{q} \;, ~~~~ \bar{\L}_U^{\rm near} \simeq \frac{m}{2-\frac{16}{3\pi}}\ln\frac{m}{\sqrt{\k}} \;,
\ee
where $a$ is given in \cref{q_a} and $y_c$ solves the equation $y_c \mathcal{H}(y_c)=a\pi/2$.
The solution always exists for our values of the dilaton barrier coupling, $a<1$, 
due to the asymptotics of the function $\mathcal{H}(y)$.

The calculation of the tunneling suppression using \cref{B_gen,Core_near,b_near} along the complex-time contour $\C$ leads to the following result (see Appendix~\ref{app:action} for the derivation),
\begin{equation}\label{B_tunn_near}
    B_{\near}^{\rm tunn.}=\frac{16\pi}{\gc^2}\left(\ln{\sqrt{\frac{\lambda'\lambda}{\kappa}}}-\frac{\lambda'}{m+q\lambda}-\frac{\lambda-\lambda'}{m\pi}\mathcal{H}(\frac{q\lambda}{m})+\ln{2}-1\right) \;.
\end{equation}
Taking $\l=\l'$ in this expression, one recovers the Hartle-Hawking vacuum decay suppression in the near-horizon region. On the other hand, setting $\l'\mapsto m$ and disregarding the order-one terms, one obtains the leading-log and the linear in $\l/m$ terms in the Unruh vacuum decay suppression \cite{Shkerin:2021rhy}.
At fixed $\l'<\l$, the suppression is a monotonically decreasing function of $\l$ subject to the condition $\bar{b}_{\near}(\l,\l')<1$.

Comparing \cref{B_tunn_far,B_tunn_near}, we see that for $\l'\lesssim \l$ the near-horizon tunneling is always dominant over the far-from-horizon decays. This includes the Hartle-Hawking and Unruh vacua and agrees with the results of Ref.~\cite{Shkerin:2021rhy} and with the expected catalyzing effect of the curved geometry of the BH. On the other hand, in the limit of cold BH, $\l'/\l\gg 1$, we find that the near-horizon tunneling is suppressed compared to the far-from-horizon decays.  
In the next section we compare the different decay channels in more detail.

\section{Decay via non-thermal activation}
\label{sec:sph}

The bounce solutions constructed in Sec.~\ref{sec:tunn} are applicable 
when the BH and/or environmental temperatures lie below the critical lines, see Fig.~\ref{fig:crit_far}. When the line is approached from below, the bounce ceases to exist.
This indicates the change in the decay regime: quantum tunneling is replaced by classical stochastic jumps of the field over the barrier separating the vacua \cite{Shkerin:2021zbf}.
What is the relevant semiclassical solution in this regime?
In the equilibrium situation, it is known as the sphaleron or the critical bubble, and the corresponding exponential suppression reproduces the classical Boltzmann formula. In our case, too, the decay is expected to occur classically, but the radiation triggering it is no longer thermal. The task is to construct the solution describing such non-thermal activation.

As we saw above, when the critical temperature line $\bar{b}_{\rm near}(\l,\l')=1$ is approached for the near-horizon bounce, the core of the bounce stops fitting the Rindler region $x<0$.
The nonlinear part of the semiclassical solution replacing eq.~(\ref{Core_near}) above the critical line is then sensitive to the full barrier potential $U_{\rm eff}(x)$, which impedes its analytic treatment in the general case $\l'\neq \l$. For the model at hand, the analysis of the Hartle-Hawking vacuum decay carried in \cite{Shkerin:2021rhy} shows that the thermal sphaleron stays in the barrier region $x\approx 0$ in a certain range of temperatures, but eventually moves to the asymptotically-flat region, where its suppression coincides with the flat-space one.\footnote{This behavior is analogous to the $4d$ Schwarzschild BH and requires the dilaton barrier be sufficiently weak, $a<1$.} 
A less accurate estimation of the decay suppression using stochastic field variances lead to the same conclusion for the Unruh vacuum, namely, that at high temperature the decay happens outside the Rindler region. 
For this reason, we first discuss decay via activation far away from the BH where we manage to construct analytically the non-thermal sphaleron in the regime $q\l/m\ll 1$. We then use the stochastic method to estimate the decay suppression in the BH vicinity and at arbitrarily high BH temperatures, $q\l/m\gtrsim 1$, thus obtaining the full picture.

\begin{figure}[t]
    \centering
    \includegraphics[width=0.45\linewidth]{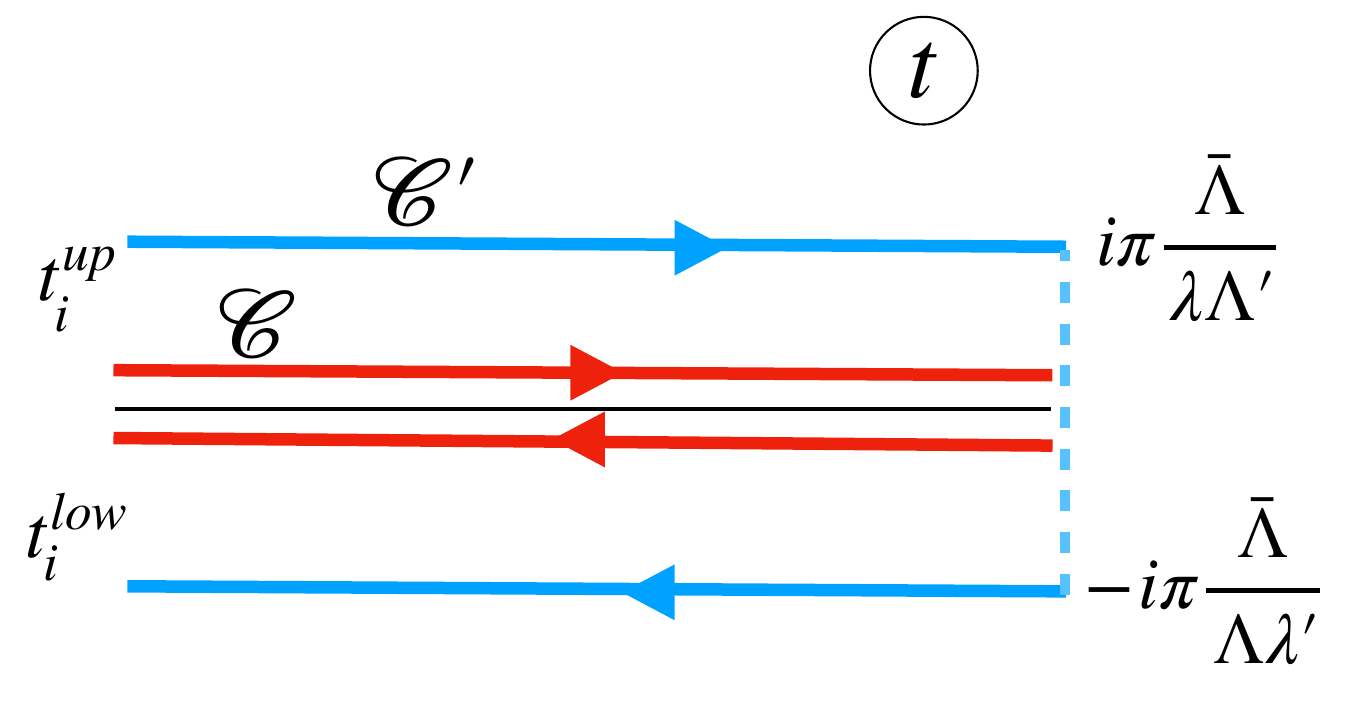}
    \caption{Contours in the complex time plane for the
      calculation of the probability of vacuum decay via activation.}
    \label{fig:contour2}
\end{figure}

\subsection{Flying sphalerons}
\label{ssec:sphalerons}

To find the sphaleron solution far away from the BH, we start by examining the far-from-horizon bounce.
The bounce core (\ref{Core_far}) at $b_{\rm far}<1$ describes the real-time field evolution after the nucleation at $t=t_f=0$. It diverges at a finite $t>0$, indicating the approach to the true vacuum at $\vf\to\infty$. When $b_{\rm far}\to 1$, the real-time singularities of the bounce are moved away from the turning point $t=t_f$, see \cref{Bounce_sings}. 
In the limit $b_{\far}=1$ the evolution towards the true vacuum is absent: the bounce becomes the ``flying sphaleron''. In the coordinates (\ref{CoordsRot}) comoving with the bounce core, the core is simply given by 
\be \label{Sphaleron}
\bb|_{\text{core}} \xrightarrow[]{b_{\rm far}\to 1} \ln \left\lbrace \frac{\l\l'}{\k  \ch^2[ \bl (\hat{x}-x_c)]}  \right\rbrace  \;, ~~~ \l,\l'\to\L,\L' \;.
\ee
However, also the evolution towards the false vacuum is absent, and the solution does not satisfy the boundary conditions in the asymptotic past.
The same problem is encountered with the static, thermal sphaleron in the equilibrium case.
The remedy is to shift the coordinates so that the left singularity is kept at a finite distance.
Namely, we do the replacement
\be 
\hat{t}\mapsto\hat{t}+\frac{1}{2\bl}\ln(1-b_{\rm far}) \;,
\ee
where $\hat{t}$ is defined in (\ref{CoordsRot}).
Note that we need to shift both $t$ and $x$ in order to follow the evolution of the moving core. 
The solution (\ref{Core_far}) becomes
\be\label{Core_far_shifted}
\bb|_{\rm core}=\ln\left\{ \frac{\lambda\lambda'b_{\rm far}}{\k[\ch(\bl x-\dl t)-\frac{1}{2}\e^{\dl x-\bl t}-\frac{1}{2}(1-b_{\rm far})\e^{-\dl x+\bl t} ]^2} \right\}\ \;,
\ee
where we also replaced $x-x_c\mapsto x$ to simplify the expressions. This can now be matched to the tail (\ref{Tail_far}) in the region $\Re t<0$, $\bl|t|\gg 1$, $|x|,|t|\ll m^{-1}$ and for any $b_{\rm far}\leqslant 1$.
At the critical line $(\l,\l')=(\L,\L')$, $b_{\rm far}=1$ and the solution reads as follows,
\be \label{Bounce_crit}
\bb^{(\L,\L')} =  \ln\left\{ \frac{\Lambda\Lambda'}{\k[\ch(\Bl x-\Dl t)-\frac{1}{2}\e^{\Dl x-\Bl t}]^2} \right\} \;,
\ee
where we denoted $\Bl\equiv(\L+\L')/2$, $\Dl\equiv(\L-\L')/2$.
The solution is linearized in the limit $\Re t\to-\infty$, and in this limit it obeys the vacuum boundary conditions \eqref{boundarycondition}. 
In other words, the core (\ref{Bounce_crit}) matches with the Green's function $\G_{\L,\L'}$ in the matching region.
On the other hand, at $\Re t\to\infty$ it approaches the flying sphaleron (\ref{Sphaleron}). 
Thus, \cref{Bounce_crit} describes the tunneling or ``jump'' onto the flying sphaleron. 
The latter decays into the true vacuum with order-one probability, hence all exponential suppression comes from
the sphaleron formation stage of the process.

\begin{figure}[t]
    \centering
    \includegraphics[width=0.3\linewidth]{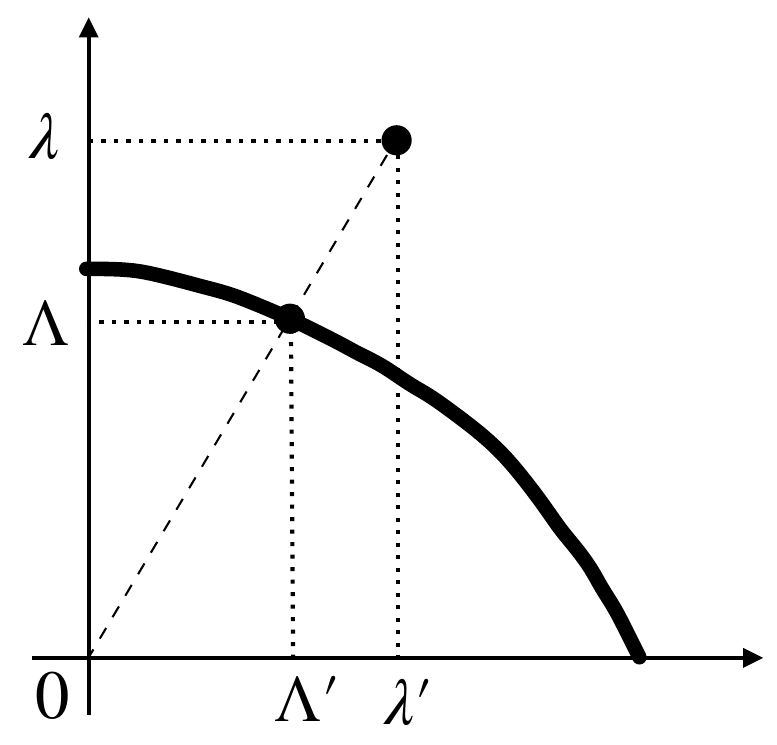}
    \caption{Choice of the point $(\L,\L')$ on the critical line for the construction of the bounce that describes decays via activation at $\l>\L$, $\l'>\L'$.}
    \label{fig:crit_far2}
\end{figure}

What is the structure of the bounce above the critical line, $\l>\L$, $\l'>\L'$? We still expect the bounce to interpolate between the false vacuum and the flying sphaleron. 
However, at these temperatures
there is no single analytic function
that would be a solution of the field equations, 
satisfy the vacuum boundary conditions at $\Re
t\to -\infty$ along the contour $\C$ and approach the sphaleron at $\Re t\to +\infty$.
In fact, the bounce cannot be found on a
contour $\C$ with a finite turn-around point $t_f$. However, we
can still construct the solution if we pull the turn-around point to
infinity as illustrated in Fig.~\ref{fig:contour2}. In this case the solution on
the upper and lower halves of the contour need not be the same
analytic function, the only requirement being that they have the same (real)
limit at $\Re t\to+\infty$.
We did not manage to find the solution in the regime when the dilaton barrier is high, $q\l/m\gtrsim 1$. However, in the regime $q\l/m\ll 1$ we find
that the following solution based on \cref{Bounce_crit} satisfies all necessary conditions:
\begin{align}
\bb^{\rm up}(t,x) &  = \bb^{(\L,\L')}\left(t+i\pi \left[ \frac{\Bl}{\L\L'} - \frac{\bl}{\l\l'} \right] , \: x+i\pi \left[ \frac{\Dl}{\L\L'} - \frac{\dl}{\l\l'} \right] \right) \;, \label{Phi_UP}\\
\bb^{\rm low}(t,x) & = \bb^{(\L,\L')}\left(t-i\pi \left[ \frac{\Bl}{\L\L'} - \frac{\bl}{\l\l'} \right] , \: x-i\pi \left[ \frac{\Dl}{\L\L'} - \frac{\dl}{\l\l'} \right] \right) \;, \label{Phi_LOW} \\
\frac{\L}{\l}&=\frac{\L'}{\l'}\ \;.\label{CondOnTemp}
\end{align}
Note the relation \eqref{CondOnTemp} between $\L$ and $\L'$ telling us which point on the critical line one should choose for the construction of the bounce. This point is uniquely determined by the ratio $\l/\l'$, see Fig.~\ref{fig:crit_far2} for illustration.

To find the decay suppression, one needs to substitute \cref{Phi_UP,Phi_LOW,CondOnTemp} to \cref{B_gen} and perform the integration. 
The task is considerably simplified by shifting the spatial variable according to (\ref{CoordsRot}) with $(\l,\l')=(\L,\L')$. The solution on the two parts of the contour becomes
\begin{align}
    \bb^{\rm up}(t,\hat{x})&=\ln \left\{ \frac{\L\L'}{\kappa\left[\ch(\bar{\L}\hat{x})-\frac{1}{2}\exp(\Dl \hat{x}-\frac{\L\L't}{\bar{\L}}+\frac{i\pi}{2}\frac{\L}{\l}+\frac{i\pi}{2}\frac{\L'}{\l'}-i\pi) \right]^2} \right\} \ \label{Phi_UP2} ,\\
    \bb^{\rm low}(t,\hat{x})&=\ln \left\{  \frac{\L\L'}{\kappa\left[\ch(\bar{\L}\hat{x})-\frac{1}{2}\exp(\Dl \hat{x}-\frac{\L\L't}{\bar{\L}}-\frac{i\pi}{2}\frac{\L}{\l}-\frac{i\pi}{2}\frac{\L'}{\l'} +i\pi) \right]^2} \right\} \ . \label{Phi_LOW2}
\end{align}
It is now manifestly periodic in the imaginary time. This allows us to deform the contour $\C$ into the new contour $\C'$, as shown in Fig.~\ref{fig:contour2}, such that the 
integrals over
the upper and lower halves of $\C'$ cancel due to the periodicity, and only the segment at $\Re t\to\infty$ contributes to the integral.
On this segment, the bounce takes the form (\ref{Sphaleron}), and the integration becomes simple. 
We obtain 
\be \label{B_sph_far}
B_{\rm sph} = \frac{2\pi}{\gc^2}\left( \frac{\L}{\l} + \frac{\L'}{\l'} \right) \left( \ln \left[ \frac{4\L\L'}{\k} \right] -4 \right) \;, ~~~ q\l/m\ll 1 \;.
\ee
This is an exact result, up to factors $\gc^{-2}\times o(1)$. It is written in the symmetric form with respect to the temperatures of the BH and the environment. The symmetry and the apparent simplicity of the expression (\ref{B_sph_far}) will go once $\L$ and $\L'$ are evaluated. We stress that $\L,\L'$ are uniquely fixed by their definition as points on the critical line and by
the condition (\ref{CondOnTemp}).
In general, they cannot be found explicitly, but we can evaluate them to the leading-log accuracy and obtain
\be\label{B_sph_far2} 
B_{\rm{sph, leading-log}} = \frac{6\pi^2m}{\gc^2\l}\frac{\left( \ln\frac{m}{\sqrt{\k}} \right)^2}{1+(\frac{3\pi}{4}-1)\frac{\l'}{\l}} \;, ~~~ q\l/m\ll 1\;.
\ee
It is straightforward to check that in the equilibrium case with $\l'=\l$, the expression (\ref{B_sph_far2}) reproduces the result of Ref.~\cite{Shkerin:2021rhy} for the suppression of the Hartle-Hawking vacuum decay far from the BH, while in the limit of the Unruh vacuum, $\l'\mapsto m$, we recover the result obtained in \cite{Shkerin:2021zbf} using the method of stochastic field variances.

The flying sphaleron solution (\ref{Phi_UP2}), (\ref{Phi_LOW2}) is not valid at very high BH temperatures, $q\l/m\gtrsim 1$, or near the BH where the spacetime is not flat. To obtain the full picture of vacuum decay, we will use the less accurate but more general method of stochastic jumps.

\subsection{Stochastic jumps}
\label{ssec:jumps}

Here we provide an alternative way to estimate the decay suppression in the high-temperature regime, $\l\geqslant\L$, $\l'\geqslant \L'$, based on the simple stochastic picture. In this regime, the occupation numbers of long modes with $\o\sim m$ are large, and field fluctuations with wavelengths $\sim m^{-1}$ are essentially classical. These non-thermal (for $\l\neq\l'$) fluctuations can throw the field over the barrier separating the vacua and trigger the decay. The rate of such events can be estimated as 
\be \label{StochRate}
\Gamma_{\rm{high}-(\l,\l')} \sim \exp\left(- \frac{\vf_{\rm max}^2 }{2(\delta\vf)_{\l,\l'}^2} \right) \;,
\ee
where $\vf_{\max}$ is the field value (\ref{Phimax}) at the maximum of the potential, and $(\delta\vf)_{\l,\l'}$ is the variance of fluctuations in the state (\ref{state}). The expression (\ref{StochRate}) assumes that the field fluctuations are Gaussian. This is a good approximation for our model, since the scalar potential (\ref{V}) with the condition (\ref{hierarchy}) is almost quadratic at $\vf<\vf_{\rm max}$. 

To find the field variance, we take the Green’s function $\G_{\l,\l'}(t,x';0,x)$ at coincident points, $t\to0,\,x'\to x$. This must be renormalized by subtracting the Feynman Green's function $\G_F$ in the Minkowski vacuum, and we obtain 
\be\label{fieldvar}
(\delta\vf)_{\l,\l'}^2={\rm g}^2\lim_{t\to 0,\,x'\to
  x}\big[\G_{\l,\l'}(t,x';0,x)-\G_{F}(t,x';0,x)\big]\;.
\ee 
Note that we can evaluate the variance close to the dilaton barrier, at $x\approx 0$, where the tunneling solution cannot be constructed analytically. As discussed in the previous section, the near-horizon bounce ceases to exist at the same time as it reaches outside of the Rindler region of the BH. By evaluating $(\delta\vf)_{\l,\l'}^2$ at $\l\geqslant\L$, $\l'\geqslant \L'$ both close to the dilaton barrier and far from it, we can determine the most probable nucleation site and the decay suppression at all temperatures of the BH and the environment.

Inspecting the Green's function in \cref{rightgreenfunction0,rightgreenfunction1,rightgreenfunction2}, we obtain the leading contributions to the field variances outside but close to the near-horizon region ($x\approx 0$) and far from the BH ($x\to\infty$):
\begin{align}
& (\delta\varphi)_{\l,\l'}^2\Big|_{x\approx 0}\simeq \frac{\gc^2\lambda'}{4\pi m}\frac{2m}{m+q\lambda}+\gc^2\frac{\lambda-\lambda'}{2\pi^2m}\mathcal{H}(\frac{q\lambda}{m}) \;,\label{vars1} \\
& (\delta\varphi)_{\l,\l'}^2\Big|_{x\rightarrow\infty}\simeq\frac{\gc^2\lambda'}{4\pi m}+\gc^2\frac{\lambda-\lambda'}{2\pi^2m}\tilde{\mathcal{H}}(\frac{q\lambda}{m}) \;,\label{vars2}
\end{align}
where the functions $\mathcal{H}(y)$, $\tilde{\mathcal{H}}(y)$ are defined in eqs.~\eqref{hq} and \eqref{htilde}, correspondingly. 
It is easy to check that in the limits $\l'=\l$ and $\l'\mapsto m$, these expressions reproduce the variances for the Hartle-Hawking and Unruh states, accordingly \cite{Shkerin:2021zbf,Shkerin:2021rhy}.\footnote{For the Unruh state, one must disregard the $\mathcal{O}(\gc^2)$ term and keep the leading $\mathcal{O}(\gc^2\l/m)$ term.}

\begin{figure}[t]
	\centering 
	\includegraphics[width=0.32\textwidth]{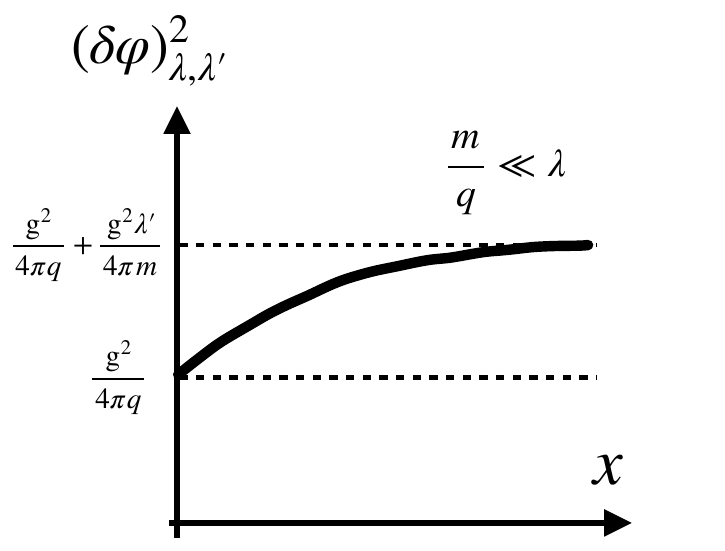}
	\includegraphics[width=0.32\textwidth]{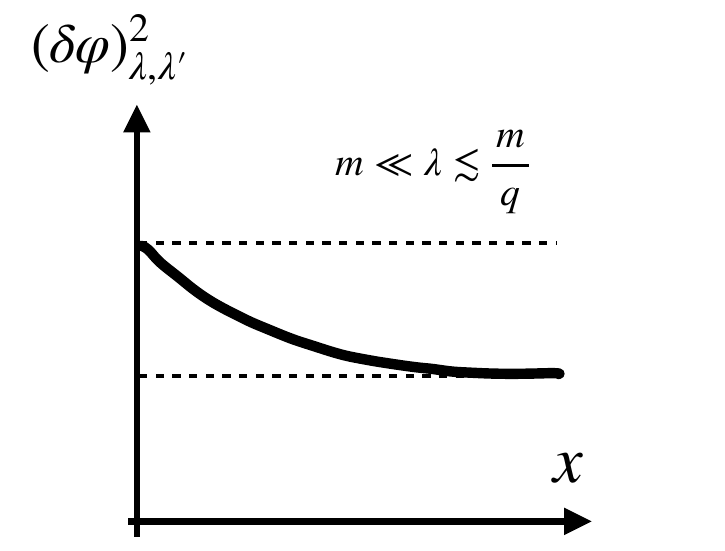}
	\includegraphics[width=0.32\textwidth]{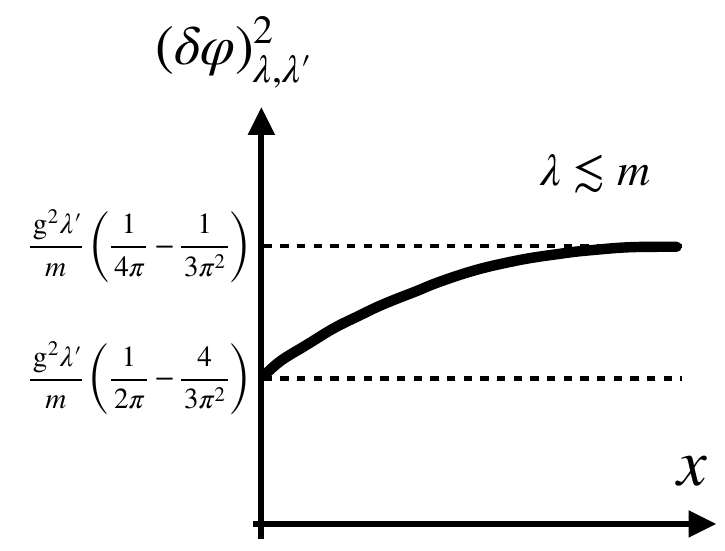}\\
(a) \qquad\qquad\qquad\qquad\qquad\qquad (b) \qquad\qquad\qquad\qquad\qquad\qquad (c)
	\caption{Variance of the field fluctuations outside the BH as a function
of the space coordinate. We take $\l'\gg m$ and the dilaton barrier $0<q < (\ln\frac{m}{\sqrt{\k}})^{-1}$. }
	\label{fig:var_far}
\end{figure}

Next, we examine \cref{vars1,vars2} in the limit of very high BH temperature, $q\l/m\gg 1$, where the functions $\mathcal{H}(y)$, $\tilde{\mathcal{H}}(y)$ are replaced by their large-$y$ asymptotics. We obtain
\be \label{vars_highT}
    (\delta\varphi)_{\l,\l'}^2\Big|_{x\approx 0}\simeq\frac{\gc^2}{4\pi q}\left(1+\frac{\lambda'}{\lambda}\right)\;,\quad 
    (\delta\varphi)_{\l,\l'}^2\Big|_{x\rightarrow\infty}\simeq  (\delta\varphi)_{\l,\l'}^2\Big|_{x\approx 0} + \frac{\gc^2 \l'}{4\pi m} \;, ~~~ q\l/m\gg 1 \;.
\ee
This behavior is illustrated in Fig.~\ref{fig:var_far}(a).
We see that at $\l'\gg m$ the variance at $x\to\infty$ is always bigger than at $x\approx 0$, meaning that the decay happens far away from the BH. This is the consequence of the dilaton barrier ``repelling'' the field fluctuations. In the Unruh vacuum limit $\l'\mapsto m$ the variance levels out at all $x>0$. The physical interpretation of this result is clear: the radiation emitted by the BH, which produces the fluctuations, remains constant at arbitrary distance from the BH due to the two-dimensional nature of the model.

\begin{figure}[t]
    \centering
    \includegraphics[width=0.7\linewidth]{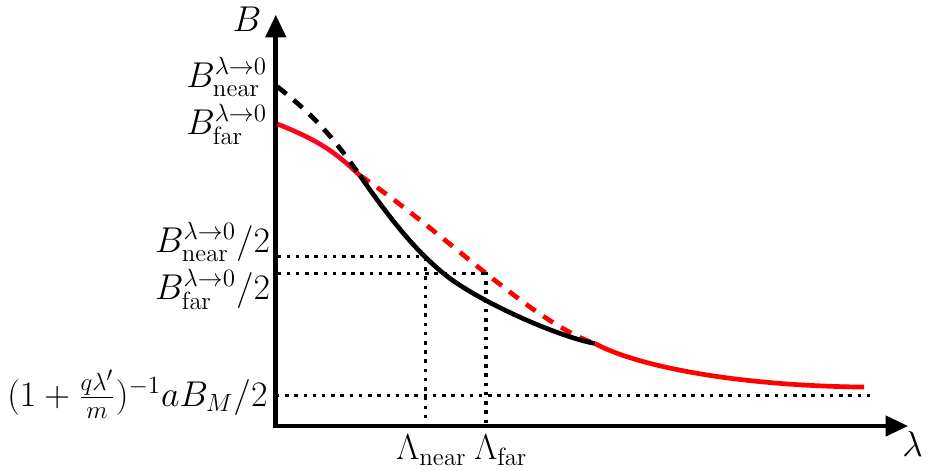}
    \caption{
    Decay suppression as a function of the BH temperature $\l$ at a fixed environment temperature $\l'<\Lambda_{HH}^{\rm near}$. 
    The black and red lines show the suppression in the BH vicinity and in the asymptotically-flat region, respectively. Solid line indicates the dominant decay channel. At $\l=\Lambda_{\rm near/far}$, the decay regime in the corresponding region changes from tunneling to stochastic jumps.}
    \label{fig:decaychannel}
\end{figure}

At not-so-large BH temperatures, $q\l/m\lesssim 1$, from \cref{vars1,vars2} 
we obtain that the decays close to the barrier, $x\approx 0$, are preferred over the decays at asymptotic infinity, see Fig.~\ref{fig:var_far}(b). Thus, keeping $\l'$ fixed and increasing $\l$, the most probable nucleation site moves from the Rindler region of the BH to the barrier region and, eventually, to the asymptotically flat region. On the other hand, 
in the limit of cold BH, $\l\mapsto m$, $\l'\gg m$, the decays always happen in the flat region (Fig.~\ref{fig:var_far}(c)). One can explain this by noticing that even when the dilaton barrier is small, $q\l\lesssim m$, a fraction of radiation approaching the BH from far away is reflected back and, hence, field fluctuations at $x>0$ are enhanced compared to the cold Rindler region $x<0$. Thus, neither very cold nor very hot BH is a good nucleation site.
This is illustrated in Fig.~\ref{fig:decaychannel} which shows the decay suppression at a fixed $\l'<\Lambda_{HH}^{\rm near}$ and varying $\l$. It shows that the near-horizon decays are only preferred in the intermediate range of BH temperatures.

\begin{figure}[t]
    \centering
    \includegraphics[width=0.4\linewidth]{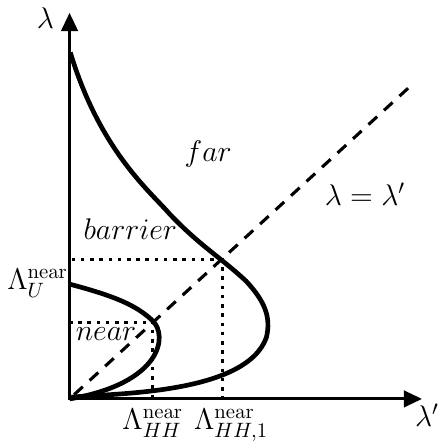}
    \caption{Regions in the parameter space $(\l,\l')$ with different nucleation sites. \textit{Near:} the tunneling happens in the BH Rindler region. \textit{Barrier}: the decay happens outside the Rindler region but close to the dilaton barrier. \textit{Far:} the decay happens in the asymptotically-flat region. 
    The dashed line is $\l=\l'$.
 }
    \label{fig:NearFar}    
\end{figure}

Figure~\ref{fig:NearFar} summarizes 
which region dominates the decays (both via tunneling and non-thermal activation) at various values of $\l,\l'$.
For $\l \gtrsim \l'$, increasing $\l$ and keeping $\l/\l'$ fixed, we start with the near-horizon tunneling, then arrive 
at the critical line ${\bar b}_\mathrm{near} = 1$, then through the barrier reach the stochastic jumps regime far from the horizon.
For example, for the Hartle-Hawking vacuum the critical line is achieved at $\l=\Lambda_{HH}^{\rm near}$ (see \cref{TempCritNear}), and the far-from-horizon regime is reached at $\l=\Lambda_{HH,1}^{\rm near}\equiv m/q$ \cite{Shkerin:2021rhy}.
For $\l \ll \l'$, the dominant decay channel is the far-from-horizon tunneling which turns to the far-from-horizon stochastic jumps upon crossing the critical line $b_\mathrm{far} = 1$.

In the limit of weak dilaton barrier, $q\l/m\ll 1$, from \cref{Phimax,StochRate,vars2} we recover
the leading-log part (\ref{B_sph_far2}) of the suppression (\ref{B_sph_far}) calculated using the flying sphaleron (\ref{Bounce_crit}).
This confirms that the decay via activation happens through the formation of the flying sphaleron, and the relevant semiclassical solution (\ref{Phi_UP}), (\ref{Phi_LOW}) describes tunneling onto this sphaleron.

Importantly, from \cref{vars_highT} we see that at $\l\to\infty$, $\l'=\rm{const}$, the field variance becomes independent of the BH temperature $\l$, implying that the exponential decay suppression never vanishes, and the rate asymptotes to
\be \label{Gamma_high_lim}
\Gamma_{\rm{high}-(\l,\l')} \to \exp\left( -\frac{aB_M}{2\left[ 1+\frac{q\l'}{m} \right]} \right) \;, ~~~ \l\to\infty \;,
\ee
where the flat-space vacuum suppression $B_M$ is given in \cref{B_M} and $a$ is defined in \cref{q_a}.
This behavior is shown in Fig.~\ref{fig:decaychannel}.
This is the consequence of the temperature-dependent dilaton barrier that cuts off the Hawking radiation at $x>0$. 
This result is expected to hold in higher dimensions as well.

\begin{figure}[t]
    \centering
    \includegraphics[width=0.6\linewidth]{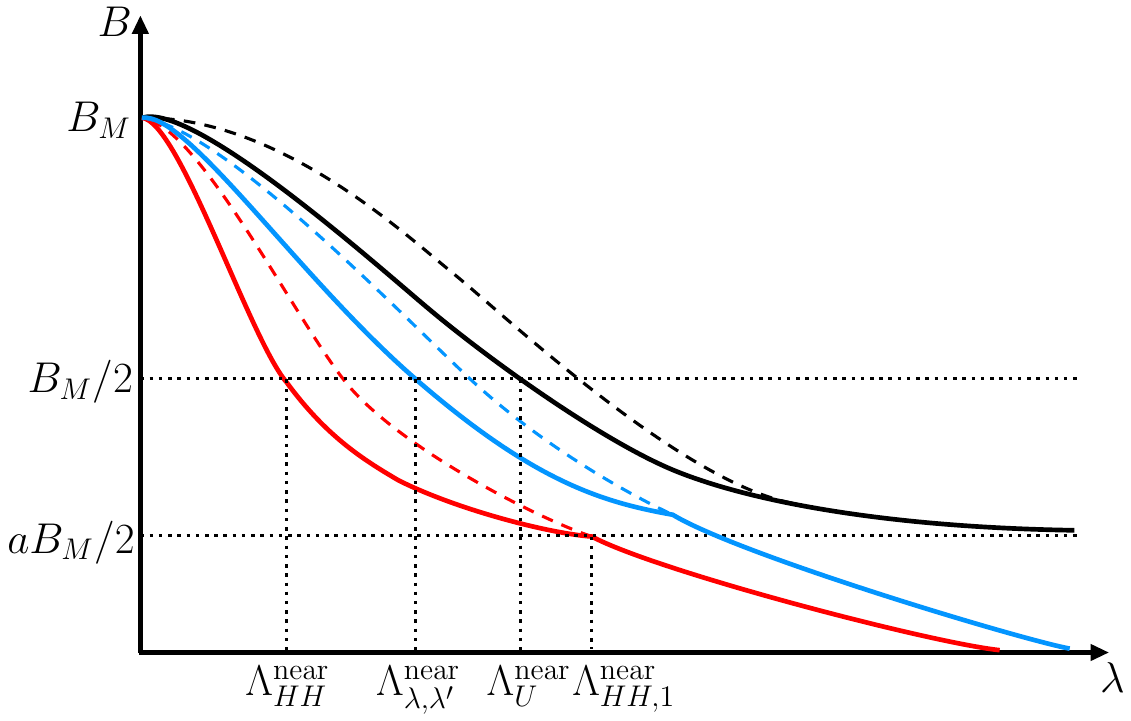}
    \caption{Decay suppression as a function of BH temperature $\l$ and different $\l'$. The black line corresponds to the Unruh vacuum with $\l'=0$,  the red line corresponds to the Hartle-Hawking vacuum with $\l'=\l$, and the blue line corresponds to an intermediate case $\l'=c\l$, $0<c<1$. The dashed lines show the decay suppression far away from the BH. At $\l=\L_{\l,\l'}^{\rm near}$, the critical line $\bar{b}_{\rm near}(\l,c\l)=1$ is reached, and 
    the decay regime in the BH vicinity changes from tunneling to stochastic jumps.
    } 
    \label{fig:B_tot}
\end{figure}

Figure~\ref{fig:B_tot} shows the decay suppression as a function of $\l$ at several values of $\l'/\l$, including the Unruh ($\l'=0$) and Hartle-Hawking ($\l'=\l$) cases. Note that for $\l'/\l>0$ the suppression vanishes in the limit of large $\l$. Interestingly, at the critical line, where the transition from tunneling to stochastic activation happens, the suppression is found to be equal to $B_M/2$ for any $\l'/\l$.\footnote{We saw this numerically, since the critical line equation cannot be solved analytically for a general $\l'/\l$.}

\section{Discussion and Conclusion}
\label{sec:disc}

In this paper we studied a toy model of vacuum decay catalyzed by a black hole (BH) 
in the presence of external heat bath with a different temperature.
To model the
BH background, we used the theory of two-dimensional dilaton gravity.
In this background, we considered the massive scalar field with the unstable Liouville potential. 
We also added the BH temperature-dependent scalar-dilaton coupling which allows us to emulate the 
four-dimensional centrifugal barrier for massive scalar modes.
Our study generalizes the results of Refs.~\cite{Shkerin:2021zbf,Shkerin:2021rhy} concerning the decay from the Hartle-Hawking and Unruh vacua, and reduces to them in the corresponding limits.

We first considered the decays via tunneling, which is relevant when the temperature of the BH and/or environment is sufficiently low. We constructed the tunneling solution corresponding to the non-equilibrium state, both in the near-horizon region and far from the BH.
We showed how the solution and the corresponding decay suppression interpolate between the thermal (Hartle-Hawking) and Unruh cases.
We also found the critical values of temperatures, at which the bounce becomes the ``flying sphaleron'' and the decay regime changes.
Above the critical line, decays proceed via sphaleron-driven activation.
Remarkably, the semiclassical solution describing the decay via (non-thermal) activation can also be constructed analytically, at least when the dilaton barrier is weak. 
The complex solution interpolates between the false vacuum in the remote past and the ``flying sphaleron'' in the asymptotic future. 
It shows that the semiclassical method based on the in-in representation of the transition amplitude is applicable in this regime, even though the relevant solution can be hard to find explicitly, as in the case of strong dilaton barrier.

We found that the exponential suppression of decay does not vanish in the limit of high BH temperature $T_{BH}$, if the temperature of the environment remains fixed. The reason is that 
at high $T_{BH}$ the decay happens outside the BH vicinity, where the flux of Hawking radiation is reduced by the barrier.
This feature has already been observed in the case of Unruh vacuum~\cite{Shkerin:2021rhy}, and our results extend it to the general non-equilibrium state (\ref{state}).
In the opposite limit of very cold (compared to the environment) BH the decay happens in the asymptotically-flat region as well. Thus, the vicinity of neither very hot nor very cold BH is a good nucleation site.
Figure~\ref{fig:NearFar} shows the most probable nucleation region as a function of the BH and environmental temperatures $\l$ and $\l'$.

Our results provide new insights into the false vacuum decay from non-equilibrium initial states, such as the state of a small, primordial evaporating BH in the early-universe plasma.
To accurately determine the catalyzing effect of such BHs, our analysis needs to be extended in several ways. First, it was performed in the two-dimensional toy model because of its simplicity that made possible the analytical treatment of the problem. 
While in this model the non-minimal coupling between the dilaton and the tunneling field generates the effective potential for field modes, which is similar to that in the four-dimensional Schwarzschild background, the dilution of the flux of particles emitted by the BH as they move away from the horizon~\cite{Candelas:1980zt,Mukaida:2017bgd} is not taken into account. 
The absence of the dilution leads to the enhancement of the decay rate at an arbitrary distance from the horizon. Thus, the four-dimensional decay rate is expected to be further suppressed once the dilution of Hawking radiation is taken into account \cite{Gorbunov:2017fhq}.

Second, we worked in the static background geometry, assuming no back-reaction of the decay to the BH. 
A notable feature of the reported enhancement of the decay rate by BHs~\cite{Gregory:2013hja,Burda:2015yfa,Burda:2016mou} is the relation of the exponential decay suppression to the change in the BH entropy associated with the change in the BH mass~\cite{Mukaida:2017bgd}. 
In the two-dimensional model studied here this relation is absent.
Thus, a consistent treatment of the problem of BH catalysis of vacuum decay must account for the dilution of the flux of Hawking radiation and the change in the BH entropy. We leave this for future study.

\section*{Acknowledgments} 

We thank Sergey Sibiryakov for useful discussions.
The work of K.K. was supported by the National Natural Science Foundation of China (NSFC) under Grant No.~W2532007 and 12547104, and by JSPS KAKENHI Grant-in-Aid for Challenging Research (Exploratory) JP23K17687.
Research at Perimeter Institute is supported in part by the Government
of Canada through the Department of Innovation, Science and Economic
Development Canada and by the Province of Ontario through the Ministry
of Colleges and Universities.

\appendix
\numberwithin{equation}{section}

\section{Dilaton black holes}
 \label{app:dilaton}

Here we outline the model of two-dimensional dilaton BH that provides the background for the tunneling field 
eq.~\eqref{S}.
The action of the model reads as follows \cite{Callan:1992rs},
\begin{equation}
    S_{\mathrm{DG}}=\int \mathrm{d}^2x \sqrt{-g}\ \e^{-2\phi}\left(R+4(\nabla_\mu\phi)^2+4\lambda^2\right)\ ,
\end{equation}
where $R$ is the scalar curvature and $\lambda$ is a constant parameter. The theory admits a one-parameter family of BH solutions with the line element
\begin{equation}
    \mathrm{d}s^2=-\Omega(r)\mathrm{d}t^2+\frac{1}{\Omega(r)}\mathrm{d}r^2\ ,
\end{equation}
where $\Omega$ and dilaton field $\phi$ are
\begin{equation}
    \Omega(r)=1-\frac{M}{2\lambda}\e^{-2\lambda r},\quad\phi=-\lambda r\ ,
\end{equation}
and $M$ is identified as the BH mass. The BH temperature is $T_{BH}=\l/(2\pi)$ and is independent of mass. The horizon radius is evaluated as
\begin{equation}
    r_h=\frac{1}{2\lambda}\ln\frac{M}{2\lambda}\ .
\end{equation}
Introducing the tortoise coordinate
\begin{equation}
    x=\frac{1}{2\lambda}\ln\left(\e^{2\lambda r}-\e^{2\lambda r_h}\right)-r_h\ ,
\end{equation}
and re-expressing $\Omega$ and $\phi$ as functions of $x$, we arrive at \cref{BH_backgr}. More properties can be referred to Appendix A in \cite{Shkerin:2021zbf}.

\section{Linear modes and Green's functions}
\label{app:Green}

Here we gather important formulas related to the effective potential \eqref{EqModes} and massive scalar linear modes in this potential. 
We then compute the Green's function $\G_{\l,\l'}$ (eq.~(\ref{GreenGen})) of the non-equilibrium state (\ref{state}) in the near-horizon region and outside the BH vicinity. 
More details on the effective potential and the calculation of the Hartle-Hawking and Unruh Green's functions, to which $\G_{\l,\l'}$ reduces in the respecting limits, can be found in Appendix C of \cite{Shkerin:2021rhy}.

\subsection{Effective potential, modes and scattering coefficients}
\label{ssec:EffPot}

The effective potential in \cref{EqModes} reads as follows, 
\begin{equation}
    U_{\eff}(x)=\frac{m^2}{1+\e^{-2\l x}}+\frac{2q\l^2\e^{-2\l x}}{(1+\e^{-2\l x})^2}\ .
\end{equation}
 When $m^2<2q\l^2$, the maximum of the barrier and its position are expressed as 
\begin{equation}
    U_{\max}=\frac{(2q\l^2+m^2)^2}{8q\l^2}\ , \quad x_{\max}=\frac{1}{2\l}\ln\left[ \frac{2q\l^2+m^2}{2q\l^2-m^2} \right]\ .
\end{equation}
In this region, the maximum of the barrier exceeds $m^2$, which is the asymptotics of the first term at positive $x$. The thickness of the barrier is of the order of $\l^{-1}$.

An advantage of this potential is that eq.~\eqref{EqModes} is exactly solvable in terms of the hypergeometric functions.
As discussed in Sec.~\ref{ssec:instate}, we take the basis of orthogonal and delta-function normalizable solutions $f_{L,\omega}$, $f_{R,\omega}$. 
The asymptotic behavior of the right-moving modes at $\o>m$ is as follows,
\begin{equation}\label{rightmoving modes}
	f_{R,\omega}=\begin{cases}
		\e^{i\omega x}+\beta_{\omega}\e^{-i\omega x}\ ,  & x\rightarrow -\infty \\
		\gamma_\omega \e^{ikx}\ , &x\rightarrow +\infty
	\end{cases}
\end{equation}
where the reflection $\beta_{\omega}$ and transmission $\gamma_{\omega}$ coefficients obey the unitarity relation
\begin{equation}\label{unitarity constraint}
	\vert\beta_{\omega}\vert^2+\frac{k}{\omega}\vert \gamma_\omega\vert^2=1 ~.
\end{equation}
Similarly, the reflection and transmission coefficients for the left-moving modes $f_{L,\omega}$ are introduced, but they are determined by $\beta_\omega$ and $\gamma_\omega$. At $0<\omega<m$, we set $f_{L,\o}\equiv0$ and define $f_{R,\o}$ as an exponentially decaying solution at $x\to\infty$, $f_{R,\o}\propto \e^{-\vk x}$ where $\vk\equiv\sqrt{m^2-\o^2}$.
The amplitudes $\beta_\o$ and $\gamma_{\o}$ are still defined by \cref{rightmoving modes} (where one should replace $k\mapsto i\vk$), and the relation \eqref{unitarity constraint} is replaced by $\vert\beta_\o\vert=1$.

When calculating the Green's function, we only need to use the expressions for $\b_\o$ and $\g_\o$ in the limit $\omega\ll\l$, where $U_{\rm eff}$ is approximated by the superposition of the step-function and the $\delta$-function. In this approximation, at $\o>m$ we find
\begin{subequations}\label{amplitudes}
\begin{align}
	\b_\o&=\frac{i(\o-k)+q\l}{i(\o+k)-q\l}~, \label{reflection}\\
	\g_\o&=\frac{2i\o}{i(\o+k)-q\l} ~, \label{transmission}
\end{align}
\end{subequations}
where we assumed $q\ll1$.  These expressions can also be used at $0<\o<m$ upon substituting $k\mapsto i\vk$.

\subsection{Green's functions in the asymptotic regions}

Here we summarize the expressions for the Green's function which are used in the main text. The derivation is postponed to the next subsection. The Green's function is evaluated in the near-horizon ({\it left}) and far-from-horizon ({\it right}) regions.
They are defined as
\begin{align}
	&x,x'<0\ ,  \qquad |x|,|x'| \gg \lambda^{-1} \qquad({\text{``left''}})\ , \label{leftcondition} \\
	&x,x'>0\ ,  \qquad x,x'\gg \l^{-1} \qquad ~~~ (\text{``right''})\ . \label{rightcondition}
\end{align}
We also assume that $m\vert x-x'\vert,m|t|\ll 1$ (``close separation'' limit).
Furthermore,
we require $m\ll\l$ and $q\ll1$, while no relation is assumed between $m$ and $q\l'$ or $q\l$. 

\subsubsection*{Green's function on the left:}

For $\l'\gg m$:
\begin{align}
	\G_{\l,\l'}|_{\rm left}^{\rm close}&=-\frac{1}{4\pi}\ln{\left[4\sh{\left(\frac{\l}{2}(x-x'-t)\right)}\sh{\left(\frac{\l'}{2}(x-x'+t)\right)}+i\epsilon\right]}-\frac{\l(x+x')}{4\pi} \nn\\
	&\quad+\frac{\l-\l'}{2\pi^2m}\mathcal{H}\left(\frac{q\lambda}{m}\right)+\frac{\lambda'}{2\pi(m+q\lambda)} \ ,  \label{leftgreenfunction1} 
\end{align}
and for $\l'\ll m$:
\begin{align} 
	\G_{\l,\l'}|_{\rm left}^{\rm close}&=-\frac{1}{4\pi}\ln\left[2\sh{\left(\frac{\l}{2}(x-x'+t)\right)}m(x-x'+t)+i\epsilon\right]-\frac{\l(x+x')}{4\pi} \nn\\
	&\quad +\frac{\l}{2\pi^2m}\H{\left(\frac{q\l}{m}\right)}+\frac{1}{8\pi}\H^{(1)}{\left(\frac{q\l}{m}\right)}+\frac{m^2\cos[m(x-x'+t)]}{2\pi^2[m^2+(q\l)^2]}\left( \frac{\l'}{m} \right)^{\frac{3}{2}}\e^{-2\pi m/\l'}\ , \label{leftgreenfunction2}
\end{align}
where 
\begin{align}
    \H(y)&=-\frac{1}{y^2}-\frac{(1+y^2)^2\arctg{y}}{y^3(y^2-1)}+\frac{\pi y}{y^2-1}\ , \label{hq}\\
    \H^{(1)}(y)&=\frac{1}{y^2}-\frac{(1+y^2)^2\ln(1+y^2)}{y^4}+2(\ln{2}-\g_E)\ .
\end{align}
These formulas are valid under the conditions \eqref{leftcondition}, $\vert x-x'+t\vert\ll \min\{m^{-1},(q\l)^{-1}\}$, and $\vert x+x'\vert>\vert t\vert$.
The functions $\H$, $\H^{(1)}$ are regular in the limit $y\rightarrow0$:
\begin{equation}
    \H(0)=\frac{8}{3},\quad \H^{(1)}(0)=-\frac{3}{2}+2(\ln2-\g_E)\ ,
\end{equation}
while in the limit $y\rightarrow\infty$ the asymptotics of $\H$ is
\begin{equation}
    \H(y)\approx\frac{\pi}{2y}\ . 
\end{equation}
The function $\H(y)$ is monotonically decreasing. 

It is easy to check that $\G_{\l,\l'}|^{\rm close}_{\rm left}$ reduces to the corresponding  expressions for the Unruh Green's function derived in \cite{Shkerin:2021rhy} in the limit $\l'\rightarrow 0$ and for the Hartle-Hawking Green's function at $\l=\l'$ in the regime $\l'\gg m$.

\subsubsection*{Green's function on the right:}
For $\l'\gg m$:
\begin{align}\label{rightgreenfunction0}
	\G_{\l,\l'}|_{\rm right}^{\rm close}=\G_{\l,\l'}|_{\rm far}^{\rm close}+\Delta\G_{\l,\l'}\ ,
\end{align}
where
\begin{align}
    \mathcal{G}_{\l,\l'}\big|_\mathrm{far}^\mathrm{close}=&-\frac{1}{4\pi}\ln{\left[4\sh{\left(\frac{\l}{2}(x-x'-t)\right)}\sh{\left(\frac{\l'}{2}(x-x'+t)\right)}+i\epsilon\right]} \nn \\
    &+\frac{\l-\l'}{2\pi^2m}\tilde{\mathcal{H}}\left(\frac{q\l}{m}\right)
	+\frac{\l'}{4\pi m} \label{rightgreenfunction1}
\end{align}
and
\begin{equation}
\label{rightgreenfunction2}
	\Delta\G_{\l,\l'}=
	\begin{cases}
		\frac{\l-\l'}{2\pi^2m}[\H(\frac{q\l}{m})-\tilde{\H}(\frac{q\l}{m})]+\frac{(\l-\l')(x+x')}{4\pi}[\frac{2q\l}{\pi m}\H(\frac{q\l}{m})-1]+\frac{\l'}{4\pi m}\frac{m-q\l}{m+q\l}\e^{-m(x+x')}, \\~~~~~~~~~~~~~~~~~~~~~~~~~~~~~~~~~~~~~~~~~~~~~~~~~~~~~~~~~~~~~~x+x'\ll\min\{\frac{1}{q\lambda},\frac{1}{m}\}\\
		\frac{\l'}{4\pi m}\frac{2m-q\l}{q\l}+\frac{\l'}{4\pi}(x+x'), ~~~~~~~~~~~~~~~~~~~~~~~~~~~~~~~~~~\frac{1}{q\lambda}\ll x+x'\ll\frac{1}{m} \\
		0, ~~~~~~~~~~~~~~~~~~~~~~~~~~~~~~~~~~~~~~~~~~~~~~~~~~~~~~~~~~~~~~~~~~~~ \frac{1}{m}\ll x+x'
	\end{cases} 
\end{equation}
Here we define
\begin{equation}\label{htilde}
    \tilde{\H}(y)=-\frac{1}{y^2}+\frac{(1+y^2)\arctg{y}}{y^3}\ .
\end{equation}
For $\l'\ll m$:
\begin{align}
   & \G_{\l,\l'}|^{\rm close}_{\rm right}=\G_{U}|_{\rm right}^{\rm close}+\sqrt{\frac{\l'}{m}}\frac{\cos{mt}}{2\pi}(1-\e^{-\frac{mxx'\l'}{\pi}})\e^{-\frac{2\pi m}{\l'}-\frac{m(x-x')^2\l'}{4\pi}} \nn \\
	&-\frac{\cos{mt}}{2\pi^2}\e^{-\frac{2\pi m}{\l'}}\frac{\sqrt{m}(x+x')\l'^{3/2}\left[ m\rm{Erfi}\left(\sqrt{\frac{m\l'}{4\pi}}\right)-q\l \right]\e^{-m(x+x')^2\l'/4\pi}-2m\l'}{(q\l)^2+m^2}\ ,  
\end{align}
where $\rm{Erfi}(y)$ is the imaginary error function.
These formulas are valid under the conditions \eqref{rightcondition}, $\vert x-x'-t\vert\ll\min\{m^{-1},(q\l)^{-1}\}$, $\vert x-x'+t\vert\ll m^{-1}$ and $x+x'>\vert t\vert$.
The function $\tilde{\H}$ is regular in the limit $y\rightarrow0$,
\begin{equation}
    \tilde{\H}(0)=\frac{2}{3}\ ,
\end{equation}
and in the limit $y\rightarrow\infty$ its asymptotics is
\begin{equation}
    \tilde{\H}(y)\approx\frac{\pi}{2y}\ .
\end{equation}
the function $\tilde{\H}(y)$ is monotonically decreasing.

The function $\G_{\l,\l'}|^{\rm close}_{\rm right}$ reduces to the corresponding expressions for the Unruh Green's function derived in \cite{Shkerin:2021rhy} in the limit $\l'\rightarrow 0$ and for the Hartle-Hawking Green's function at $\l=\l'$ in the limit $\l'\gg m$.

\subsection{Calculation of the Green's functions}

Here we provide some details of the calculation of the Green's function. First we present the general expression for $\G_{\l,\l'}$. Using \cref{state} for the occupation numbers, the direct calculation yields (we restore $t'$ in the general expressions below)
\begin{align}
\label{Green_Gen0}
	&\mathcal{G}_{\l,\l'}(t,x;t',x')
	=\int_{0}^{\infty}\frac{\mathrm{d}\omega}{4\pi\omega}\left\{
	f_{R,\omega}(x)f_{R,\omega}^*(x')\left[\frac{\e^{-i\omega\vert t-t'\vert}}{1-\e^{-2\pi\omega/\lambda}}+\frac{\e^{i\omega\vert t-t'\vert}}{\e^{2\pi\omega/\lambda}-1}\right] \right. \nonumber \\
	&\quad+f_{L,\omega}(x)f_{L,\omega}^*(x')\left[\frac{\e^{-i\omega\vert t-t'\vert}}{1-\e^{-2\pi\omega/\lambda'}}+\frac{\e^{i\omega\vert t-t'\vert}}{\e^{2\pi\omega/\lambda'}-1}\right]\nonumber\\
	&\quad+\left(\vert\beta_\omega\vert^2-1\right)\left[f_{R,\omega}(x)f^*_{R,\omega}(x')-f_{L,\omega}(x)f^*_{L,\omega}(x')\right]\left(\frac{\e^{i\omega(t-t')}}{\e^{2\pi\omega/\lambda}-1}-\frac{\e^{i\omega(t-t')}}{\e^{2\pi\omega/\lambda'}-1}\right)\nonumber\\
	&\left. \quad+\sqrt{\frac{k}{\omega}}\left[\gamma_\omega\beta_\omega^*f_{R,\omega}(x)f^*_{L,\omega}(x')+\gamma_\omega^*\beta_\omega f_{L,\omega}(x)f^*_{R,\omega}(x')\right]\left(\frac{\e^{i\omega(t-t')}}{\e^{2\pi\omega/\lambda}-1}-\frac{\e^{i\omega(t-t')}}{\e^{2\pi\omega/\lambda'}-1}\right)\right\}~, 
\end{align}
where we have used the relations between the modes and their complex conjugates (see Appendix B of \cite{Shkerin:2021zbf} for details). 
The first two lines are thermal contributions from right-moving modes and left-moving modes,  respectively. The last two lines contain the mixing terms depending on the reflection and transmission amplitudes $\b_\o$, $\g_\o$. 
Analyzing the asymptotic behavior of the integrand at $\o\rightarrow+\infty$, we conclude that $\G_{\l,\l'}$ is analytic in the strip $\vert\Im(t-t')\vert<\min\{2\pi/\l,2\pi/\l'\}$ except for usual singularities on the real axis and singularity in the amplitude $\g_\o$.

It is useful to rewrite the expression in (\ref{Green_Gen0}) as the sum of 
either Hartle-Hawking or Unruh Green's function and an additional contribution from the left-moving modes:
\begin{align}
	\mathcal{G}_{\l,\l'}&= \mathcal{G}_{HH}-\int_{0}^{\infty}\frac{\mathrm{d}\omega}{4\pi\omega}\left\{f_{L,\o}(x)f^*_{L,\o}(x')\frac{\e^{-i\omega(t-t')}}{\e^{2\pi\omega/\lambda}-1}+f_{L,\o}^*(x)f_{L,\o}(x')\frac{\e^{i\omega(t-t')}}{\e^{2\pi\omega/\lambda}-1}\right\}\nonumber \\
	&\quad+\int_{0}^{\infty}\frac{\mathrm{d}\omega}{4\pi\omega}\left\{f_{L,\o}(x)f^*_{L,\o}(x')\frac{\e^{-i\omega(t-t')}}{\e^{2\pi\omega/\lambda'}-1}+f_{L,\o}^*(x)f_{L,\o}(x')\frac{\e^{i\omega(t-t')}}{\e^{2\pi\omega/\lambda'}-1}\right\}  \\
	&=\G_U+\int_{0}^{\infty}\frac{\mathrm{d}\omega}{4\pi\omega}\left\{f_{L,\o}(x)f^*_{L,\o}(x')\frac{\e^{-i\omega(t-t')}}{\e^{2\pi\omega/\lambda'}-1}+f_{L,\o}^*(x)f_{L,\o}(x')\frac{\e^{i\omega(t-t')}}{\e^{2\pi\omega/\lambda'}-1}\right\} . \label{Greenfunctionrelation}
\end{align}
See Ref.~\cite{Shkerin:2021zbf} for explicit formulas for $\mathcal{G}_{HH}$ and $\mathcal{G}_U$. In the following, we will calculate $\G_{\l,\l'}$ in the several limiting cases.
\\

\noindent \textbf{Green's function on the left.\ }
First we calculate $\G_{\l,\l'}$ on the left. We start from the relation \eqref{Greenfunctionrelation} between the general Green's function $\G_{\l,\l'}$ and the Unruh Green's function $\G_U$. Using the expressions of $f_{L,\omega}$ at different $\o$ one finds
\begin{equation}
	\G_{\l,\l'}|_{\rm left}=\G_{U}|_{\rm left}+\int_{m}^{\infty}\frac{\mathrm{d}\omega\ k}{2\pi\omega^2}\vert\gamma_\omega\vert^2\frac{\cos{[\omega(x-x'+t)]}}{e^{2\pi\omega/\lambda'}-1} \ .
\end{equation}
The first term was calculated in \cite{Shkerin:2021rhy}. We denote the second term by $\G_{\l,\l'}^{(2)}$ and evaluate it in the two limiting cases.

We first consider the regime $\l'/m\ll1$ corresponding to the Unruh limit and work out the leading $\l'/m$ contribution. 
To simplify the integrand, we note that the exponential $\e^{2\pi\o/\l'}$ in the denominator dominates at all $\o>m$, and thus the denominator can be replaced by 
$\e^{-2\pi\o/\l'}$. Introducing the variable $s \equiv \o-m$, the integral becomes
\begin{equation}
	\G_{\l,\l'}^{(2)}=\e^{-2\pi m/\l'}\int_{0}^{\infty}\mathrm{d}s\ \frac{2\sqrt{s^2+2ms}\cos{[(m+s)(x-x'+t)]}}{\pi[(q\l)^2+(m+s+\sqrt{s^2+2ms})^2]}\e^{-2\pi s/\l'}\ .
\end{equation}
Due to the exponential suppression factor $\e^{-2\pi m/\l'}$, only the interval $s< \l'/2\pi\ll m$ contributes to $\G_{\l,\l'}^{(2)}$. Expanding the integrand in $s/m$ and keeping the leading term, we obtain 
\begin{equation}
	\G_{\l,\l'}^{(2)}=\frac{m^2\cos[m(x-x'+t)]}{2\pi^2[m^2+(q\l)^2]}\left( \frac{\l'}{m} \right)^{\frac{3}{2}}\e^{-2\pi m/\l'} \ .
\end{equation}

\begin{figure}[t]
    \centering
    \includegraphics[width=1\linewidth]{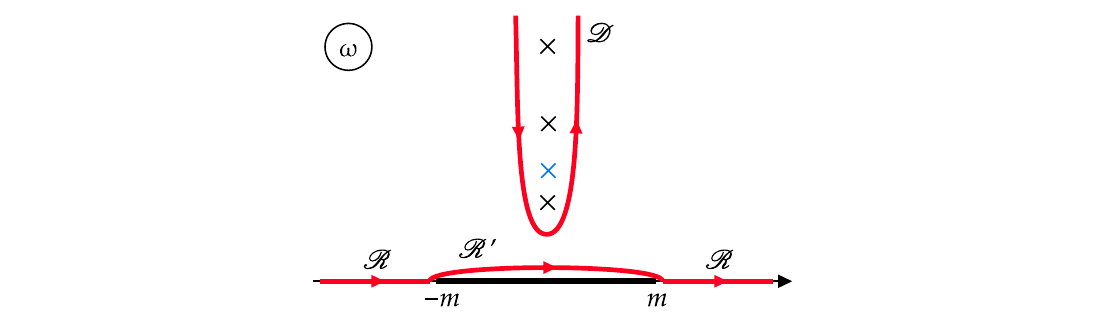}
    \caption{Contours in the $\o$-plane used in the calculation of the Green's function at $q\l>m$. The blue cross is the pole of the transmission coefficient $\vert\g_\o\vert^2$, the black crosses are thermal poles.}
    \label{AppB}
\end{figure}

The opposite limit $\l'\gg m$ proves to be more challenging.
However, in this limit the integral $\G_{\l,\l'}^{(2)}$ is similar to the one arising in the Unruh case \cite{Shkerin:2021rhy}. 
We continue analytically the functions $k(\o)$ and $\vert\g_\o\vert^2$ in the upper half-plane.  Using $k(-\omega)=-k(\omega)$, $\vert\gamma_{-\omega}\vert^2=\vert\gamma_{\omega}\vert^2$ for $\omega>m$, we arrive at
\begin{eqnarray}\label{G2}
	\G_{\l,\l'}^{(2)}=\int_{\mathscr{R}}\frac{ \mathrm{d}\omega\ k}{4\pi\omega^2}\vert\gamma_\omega\vert^2\frac{\e^{i\omega(x-x'+t)}}{1-\e^{-2\pi\omega/\lambda'}}-\int_{m}^{\infty}\frac{\mathrm{d}\omega\ k}{4\pi\omega^2}\vert\gamma_\omega\vert^2\e^{i\omega(x-x'+t)}\ ,
\end{eqnarray}
where $\mathscr{R}=(-\infty,-m]\cup[m,\infty)$. Assuming that $x-x'+t>0$ (the result will eventually be analytically continued to $x-x'+t<0$), we divide $\G_{\l,\l'}^{(2)}$ into four parts,
\begin{equation}
 \G^{(2)}_{\l,\l'}=\G^{(21)}_{\l,\l'}+\G^{(22)}_{\l,\l'}+\G^{(23)}_{\l,\l'}+\G^{(24)}_{\l,\l'}\ ,
\end{equation}
where the first three terms correspond to the first integral in \eqref{G2} evaluated along different contours as shown in Fig.~\ref{AppB}. 
Namely, we add $\mathscr{R}'$, which is the upper side of the branch cut at $-m<\o<m$, to $\mathscr{R}$. The resulting contour is deformed into $\mathscr{D}$, which encircles the poles at the imaginary axis, and this can be evaluated using the residue theorem. Thus, the integral along the contour $\mathscr{R}$ is transformed into the integral along the contour $\mathscr{D}$ subtracted by that along $\mathscr{R}'$. $\G_{\l,\l'}^{(21)}$ and $\G_{\l,\l'}^{(23)}$ are evaluated along $\mathscr{D}$, where $\G^{(21)}_{\l,\l'}$ picks up the thermal poles at $\o=in\l', n=1,2,\dots$, while $\G_{\l,\l'}^{(23)}$ picks up the pole of $\vert\g_\o\vert^2$. Next,
$\G_{\l,\l'}^{(22)}$ is evaluated along $-\mathscr{R}'$. 
Finally, $\G_{\l,\l'}^{(24)}$ is the second integral in \eqref{G2}. 

In evaluating $\mathcal{G}_{\l,\l'}^{(21)}$ we can set $\vert\gamma_\omega\vert^2=1$ since 
we assumed $m\ll\lambda'$ and $q\ll1$. 
Then the summation of residues gives
\begin{equation}
	\G_{\l,\l'}^{(21)}=-\frac{1}{4\pi}\ln{\left[1-\e^{-\lambda'(x-x'+t)}\right]}\ .
\end{equation}

To evaluate $\G_{\l,\l'}^{(22)}$, we assume $m(x-x'+t)\ll1$ and expand the integrand to the subleading orders in $\o/\l'$ and $\o(x-x'+t)$.  
The integral becomes
\begin{equation}
	\G_{\l,\l'}^{(22)}=-\frac{i\lambda'}{2\pi^2}\int_{-m+i\epsilon}^{m+i\epsilon}\frac{\mathrm{d}\o\ \vk}{\o((\o+i\vk)^2+(q\l)^2)}-\left(\frac{i}{2\pi}-\frac{\l'(x-x'+t)}{2\pi^2}\right)\int_{-m}^{+m}\frac{\mathrm{d}\omega\ \vk}{(\o+i\vk)^2+(q\lambda)^2}\ .
\end{equation}
Evaluating the integrals, we arrive at:
\begin{align}
	\mathcal{G}_{\l,\l'}^{(22)}&=\frac{\l'}{4\pi^2m}\left\{\frac{2}{y^2}+\frac{(1+y^2)^2\arctg{\left(\frac{2y}{1-y^2}\right)}}{y^3(y^2-1)}-\frac{2\pi}{y^2-1}\right\} \nn\\ 
    &\quad -\left(-\frac{i}{8}+\frac{\l'(x-x'+t)}{8\pi}\right)\left(\theta(1-y)-\frac{1+2y^2}{y^4}\theta(y-1)\right) \ ,
\end{align}
where we denoted
\begin{equation}
	y=\frac{q\l}{m} \label{y}
\end{equation}
and $\theta(x)$ is the Heaviside step-function with $\theta(0)=\frac{1}{2}$.

Before evaluating $\G_{\l,\l'}^{(23)}$, we analyze the singularity of the transmission coefficient $\vert\g_\o\vert^2$. From \cref{transmission}  we see that there are no poles on the upper half-plane for $q\l<m$. We can find a single pole for $q\l>m$ at
\begin{equation}
	\o_q=\frac{i}{2q\l}((q\l)^2-m^2)\ .
\end{equation}
We can now use its residue to calculate the integral.
Note that the pole $\o_q$ may overlap with one thermal pole in $\l'$ if we do not put an extra restriction between $\l$ and $\l'$. 
However, one can prove that it is sufficient to count the contributions from the two poles separately. In the close separation limit, 
$m(x-x'+t)\ll1$ and 
$q\l(x-x'+t)\ll1$, we thus obtain 
\begin{equation}
	\G_{\l,\l'}^{(23)}=\left[\frac{\l'}{4\pi m}\frac{(1+y^2)^2}{y^3(y^2-1)}-\left(-\frac{i}{8}+\frac{\lambda'(x-x'+t)}{8\pi}\right)\frac{(1+y^2)^2}{y^4}\right]\theta(y-1)\ .
\end{equation}
It can be seen that the discontinuities at $y=1$ appearing in $\G_{\l,\l'}^{(22)}$, $\G_{\l,\l'}^{(23)}$ and the pole at $\o=\o_q$ in the upper half-plane at $y>1$ are cancelled. The way to calculate $\mathcal{G}_{\l,\l'}^{(24)}$ is exactly the same as $\mathcal{G}^{(24)}_U$ in \cite{Shkerin:2021rhy}, 
since there is no dependence on $\lambda'$ in this integral. The result is
\begin{equation}
	\G_{\l,\l'}^{(24)}
	=\frac{1}{4\pi}\ln{[m(x-x'+t)]}-\frac{\ln2-\gamma_E}{4\pi}-\frac{i}{8}-\frac{y^2-(1+y^2)^2\ln[1+y^2]}{8\pi y^4}\ .
\end{equation}

Combining all four contributions, performing the analytic continuation to $x-x'+t<0$ and adding the result to $\G_U|_{\rm left}$, we finally arrive at \cref{leftgreenfunction1}.\\

\noindent\textbf{Green's function on the right.\ }
Next we evaluate the Green's function $\G_{\l,\l'}$ on the right. We use the expression for $f_{L,\o}$ in the asymptotic regions given in \cref{Greenfunctionrelation}. The general Green's function $\G_{\l,\l'}$ is related to the Unruh Green's function as follows,
\begin{align}
	\G_{\l,\l'}|_{\rm right}&=\G_{U}|_{\rm right}+\int_{m}^{\infty}\frac{\mathrm{d}\o}{2\pi k}\frac{\cos{[k(x-x')+\o t]}+\vert\b_\o\vert^2\cos{[k(x-x')-\o t]}}{\e^{2\pi\o/\l'}-1} \nn \\
	&\quad -\int_{m}^{\infty}\frac{\mathrm{d}\o}{2\pi k}\left[\frac{\g_{\o}^*\b_\o}{\g_{\o}}\e^{-ik(x+x')}+\frac{\g_{\o}\b_\o^*}{\g_{\o}^*}\e^{ik(x+x')}\right]\frac{\cos{\o t}}{\e^{2\pi\o/\l'}-1} \ . \label{gllpr}
\end{align}
In the limit $\l'\ll m$ we can compute the second and the third terms following the same steps as for the Green's function on the left. The result is 
\begin{align}
	\G_{\l,\l'}|_{\rm right}&=\G_{U}|_{\rm right}+\sqrt{\frac{\l'}{m}}\frac{\cos{mt}}{2\pi}\e^{-\frac{2\pi m}{\l'}-\frac{m(x-x')^2\l'}{4\pi}}(1-\e^{-\frac{mxx'\l'}{\pi}}) \nn \\
	&-\frac{\cos{mt}}{2\pi^2}\e^{-\frac{2\pi m}{\l'}}\frac{\sqrt{m}(x+x')\l'^{3/2}\left[ m\rm{Erfi}\left(\sqrt{\frac{m\l'}{4\pi}}\right)-q\l \right]\e^{-m(x+x')^2\l'/4\pi}-2m\l'}{(q\l)^2+m^2}\ , 
\end{align}
where $\rm{Erfi}(y)$ is the imaginary error function.

For the case $\l'\gg m$, the second term of eq.~\eqref{gllpr}, which we denote by $\tilde{\G}_{\l,\l'}^{(2)}$ can also be calculated by the same method as previously. Omitting the details, we just show the answer:
\begin{align}
	\tilde{\G}_{\l,\l'}^{(2)}&=\frac{1}{4\pi}\ln{\left[ \frac{m(x-x'+t)}{2\sh{\frac{\l'}{2}(x-x'+t)}}\right]}-\frac{\l'}{2\pi^2m}\left(-\frac{1}{y^2}+\frac{1+y^2}{y^3}\arctg{y}-\frac{\pi}{2}\right) \nn\\
	&\quad -\frac{1}{8\pi}\left[\frac{1}{y^2}+\frac{y^4-1}{y^4}\ln{(1+y^2)}+2(\ln{2}-\g_E)\right]\ ,
\end{align}
where $y$ is defined in \eqref{y}. To obtain this result, we used $m\vert x-x'+t\vert\ll1$, $m\vert x-x'-t\vert\ll1$ and $q\l \vert x-x'-t\vert\ll1$.

We denote the third term of eq.~\eqref{gllpr} by $\tilde{\G}_{\l,\l'}^{(3)}$.  Substituting the expressions \eqref{amplitudes} for $\g_\o$ and $\b_\o$, the integral is written as
\begin{equation}
	\tilde{\G}_{\l,\l'}^{(3)}=-\int_{m}^{\infty}\frac{\mathrm{d}\o}{2\pi k}\left[\frac{i(\o-k)+q\l}{i(\o+k)+q\l}\e^{-ik(x+x')}+\text{h.c.}\right]\frac{\cos{\o t}}{\e^{2\pi\o/\l'}-1}\ .
\end{equation}
Then, $\tilde{\G}_{\l,\l'}^{(3)}$ can be treated in the same way as $\tilde{\G}_{U}^{(3)}$ from \cite{Shkerin:2021rhy}. We expand the thermal
factor up to the first subleading term and assume $\vert t\vert<x+x'\ll\min\{m^{-1},(q\lambda)^{-1}\}$. The result is
\begin{align}
	\tilde{\G}_{\l,\l'}^{(3)}&=-\frac{\lambda'}{2\pi^2m}\left[\frac{\pi(y^2+1)}{2(y^2-1)}-\frac{2(1+y^2)}{y(y^2-1)}\arctg{y}\right]+\frac{(1+y^2)\ln{(1+y^2)}}{4\pi y^2}\nonumber\\
	&\quad -\frac{\lambda'(x+x')}{4\pi}\left[-\frac{2}{\pi y}+\frac{2y}{y^2-1}-\frac{2(1+y^2)^2}{\pi y^2(y^2-1)}\arctg{y}\right]\ .
\end{align}
If $x+x'\gg m^{-1}$, the integrand is rapidly oscillating, and hence we evaluate $\tilde{\mathcal{G}}^{(3)}\approx0$. For the intermediate range $(q\lambda)^{-1}\ll x+x'\ll m^{-1}$, which is only possible at $q\lambda\gg m$, we obtain
\begin{equation}
	\tilde{\G}_{\l,\l'}^{(3)}=-\frac{\lambda'}{2\pi^2m}\left(\frac{\pi}{2}-\frac{\pi}{y}\right)-\frac{\lambda'(x+x')}{4\pi}\left(-1+\frac{2}{y}\right)-\frac{1}{2\pi}\ln{[m(x+x')]}+\frac{\ln{2}-\gamma_E}{2\pi}\ .
\end{equation}
Adding up all the contributions to $\G_{U}|_{\rm right}$, we obtain \cref{rightgreenfunction1} and \cref{rightgreenfunction2}.

\section{Calculation of the bounce suppression}
\label{app:action}

Here we provide some details of the calculation of the bounce action presented in Sec.~\ref{sec:tunn}. 

\

\noindent\textbf{Bounce far from horizon.\ }
First we evaluate the bounce action for the tunneling far from horizon. Substituting the nonlinear core solution \eqref{Core_far} into the formula \eqref{B_gen}, it is written as
\begin{align}\label{B_far_C}
    B_{\far}=\frac{i}{\gc^2}&\int_{-\infty}^{\infty}\mathrm{d}x\int_{\mathcal{C}}\mathrm{d}t\frac{\l\l'b_{\far}}{\{ \ch[\bl(x-x_c)-\dl t]-\sqrt{1-b_{\far}}\ch[\bl t-\dl(x-x_c)] \}^2} \nn \\
    &\times\left\{ \ln\left\{ \frac{\l\l'b_{\rm far}}{\k\{ \ch[ \bl (x-x_c) - \dl t ]-\sqrt{1-b_{\rm far}}\ch[ \bl t - \dl(x-x_c) ]  \}^2} \right\}-2 \right\}\ .
\end{align}
The time integral is performed over contour $\mathcal{C}$ encircling the singularity of the integrand. The singularity is a second-order pole superimposed on a logarithmic branch cut. We use Feynman technique to deal with this integral. We first divide the integral into two parts
\begin{equation}\label{B_far_C2}
    B_{\far}=\frac{\lambda'\lambda b_{\far}}{\gc^2}\left[ 2I_1+\left(\ln\left[\frac{\lambda'\lambda b_{\far}}{\k}\right]-2\right)I_2 \right]\ ,
\end{equation}
where we denoted 
\begin{align}
    I_1 &= i\int_{-\infty}^{\infty}\mathrm{d}x\int_{\mathcal{C}}\mathrm{d}t\frac{-\ln\{(\ch[\bl(x-x_c)-\dl t]-\sqrt{1-b_{\far}}\ch[\bl t-\dl(x-x_c)]\}}{\{\ch[\bl(x-x_c)-\dl t]-\sqrt{1-b_{\far}}\ch[\bl t-\dl(x-x_c)]\}^2} \ ,\\
	I_2 &=i\int_{-\infty}^{\infty}\mathrm{d}x\int_{\mathcal{C}}\mathrm{d}t\frac{1}{\{\ch[\bl(x-x_c)-\dl t]-\sqrt{1-b_{\far}}\ch[\bl t-\dl(x-x_c)]\}^2}\ .
\end{align}
First, we evaluate $I_2$ by the residue of the second-order pole $t_0(x)$, since there is no logarithmic branch cut. We have
\begin{equation}
    I_2=2\pi\int_{-\infty}^{\infty}\mathrm{d}x\ \lim_{t\rightarrow t_0(x)}\frac{\d}{\d t}\left[ \frac{(t-t_0)^2}{\{\ch[\bl(x-x_c)-\dl t]-\sqrt{1-b_{\far}}\ch[\bl t-\dl(x-x_c)]\}^2} \right]\ . 
\end{equation}
This integral can be easily calculated upon introducing a new variable 
\begin{equation}
    w=\sh[\bl(x-x_c)-\dl t_0]\ .
\end{equation}
Noting that $\mathrm{d}t_0(x)/\mathrm{d}x$ can be derived from the equation requiring the denominator of the integrand to vanish, we obtain the final result
\begin{equation}
    I_2=\frac{4\pi}{\lambda'\lambda b_{\far}}\ .
\end{equation}

\begin{figure}[t]
    \centering
    \includegraphics[width=0.6\linewidth]{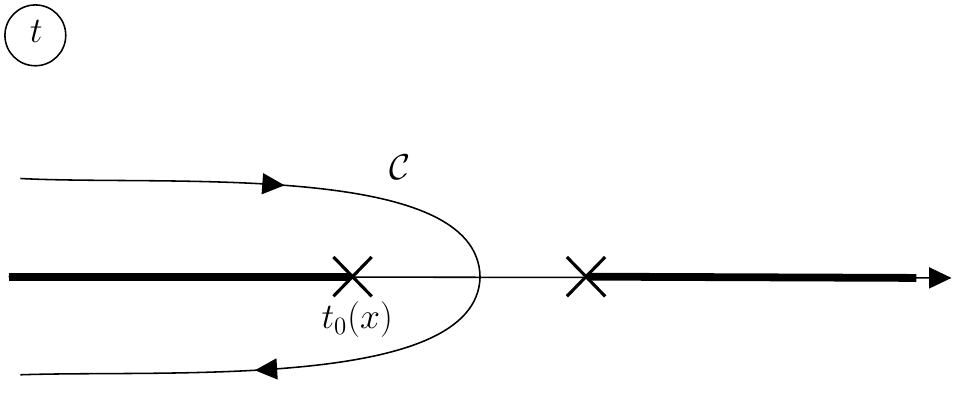}
    \caption{Time integration contour in \eqref{B_far_C}.}
    \label{timecontour}
\end{figure}

We turn to $I_1$, which is more difficult to evaluate because of the branch cut.
To avoid the branch cut, consider instead the following integral,
\begin{equation}
    I(a)=i\int_{-\infty}^{\infty}\mathrm{d}x\int_{\mathcal{C}}\mathrm{d}t\frac{-\ln\{(\ch[\bl(x-x_c)-\dl t]-a\ch[\bl t-\dl(x-x_c)]\}}{\{\ch[\bl(x-x_c)-\dl t]-\sqrt{1-b_{\far}}\ch[\bl t-\dl(x-x_c)]\}^2} \ .
\end{equation}
Using the Feynman technique, we find
\begin{equation}
    I_1=I(\sqrt{1-b_\mathrm{far}})=I(0)+\int_{0}^{\sqrt{1-b_\mathrm{far}}}I'(a)\mathrm{d}a\ .
\end{equation}
The advantage of this expression is that both $I(0)$ and $I'(a)$ are regular except for singularities along the real time axis. Thus, we can use the same method as for $I_2$ to evaluate $I_1$.\footnote{New singularities introduced from the denominator of $I'(a)$ are always encircled by the contour $\mathcal{C}$.}  We first evaluate the residue and then perform the integral over $x$ by changing the variables. Omitting the details, the result is
\begin{equation}
    I_1=\frac{4\pi}{\l'\l b_{\far}}\left(-1-\ln{\frac{b_{\far}}{2}}\right)\ .
\end{equation}
Substituting the results of $I_1$ and $I_2$ into \cref{B_far_C2} we finally obtain
\begin{equation}
    B_{\far}=\frac{4\pi}{\gc^2}\left( \ln \left[\frac{4\l\l'}{\k b_{\far}}\right]-4 \right)\ ,
\end{equation}
which yields \cref{B_tunn_far} after substituting $b_{\far}$ from \cref{b_far}.\\

\noindent\textbf{Bounce near horizon.\ }
Next we evaluate the bounce action for the tunneling near horizon. Using
the expression \eqref{Core_near}, the bounce action is written as
\begin{align}\label{B_near_C}
    B_{\near}&=\frac{i}{\gc^2}\int_{-\infty}^{\infty}\mathrm{d}x\int_{\mathcal{C}}\mathrm{d}t\frac{\l\l'b_{\near}}{\{ \ch[\bl(x-x_c)-\dl t]-\sqrt{1-b_{\near}}\ch[\bl t-\dl(x-x_c)] \}^2} \nn \\
    &\times\left\{ \ln\left\{ \frac{\l\l'b_{\rm near}\e^{-2\l(x-x_c)}}{\k\{ \ch[ \bl (x-x_c) - \dl t ]-\sqrt{1-b_{\rm near}}\ch[ \bl t - \dl(x-x_c) ]  \}^2} \right\}-2\l x_c-2 \right\}\ .
\end{align}
Fortunately, the contribution from the linear piece $-2\l(x-x_c)$ vanishes,\footnote{To see this, one can make the transformation $x\rightarrow -x$, $t\rightarrow -t$, which flips the sign of the integral, and check that the integration over the transformed contour gives the same result as before the transformation.}
and the remaining terms coincide with those appearing in $B_{\far}$.
Similarly to $B_{\far}$, we then divide $B_{\near}$ into two parts,
\begin{equation}
	B_{\near}=\frac{\lambda'\lambda b_{\near}}{\gc^2}\left[ 2I_1+\left(\ln\left[ \frac{\lambda'\lambda b_{\near}}{\kappa} \right]-2\lambda x_c-2\right)I_2 \right]\ ,
\end{equation}
where $I_1$ and $I_2$ have the same form as in the previous paragraph with $b_{\far}$ being replaced by $b_{\near}$. Using these expressions, we obtain
\begin{equation}
    B_{\near}=\frac{4\pi}{\gc^2}\left( \ln \left[\frac{4\l\l'}{\k b_{\near}} \right]-2\l x_c-4 \right). \label{bnear}
\end{equation}
After substituting \cref{b_near} to eq.~\eqref{bnear}, we arrive at \cref{B_tunn_near}.

\bibliographystyle{JHEP}
\bibliography{ref}

@article{Andreassen:2016cvx,
      author         = "Andreassen, Anders and Farhi, David and Frost, William
                        and Schwartz, Matthew D.",
      title          = "{Precision decay rate calculations in quantum field
                        theory}",
      journal        = "Phys. Rev.",
      volume         = "D95",
      year           = "2017",
      number         = "8",
      pages          = "085011",
      doi            = "10.1103/PhysRevD.95.085011",
      eprint         = "1604.06090",
      archivePrefix  = "arXiv",
      primaryClass   = "hep-th",
      SLACcitation   = "%%CITATION = ARXIV:1604.06090;%%"
}

@article{Andreassen:2016cff,
      author         = "Andreassen, Anders and Farhi, David and Frost, William
                        and Schwartz, Matthew D.",
      title          = "{Direct Approach to Quantum Tunneling}",
      journal        = "Phys. Rev. Lett.",
      volume         = "117",
      year           = "2016",
      number         = "23",
      pages          = "231601",
      doi            = "10.1103/PhysRevLett.117.231601",
      eprint         = "1602.01102",
      archivePrefix  = "arXiv",
      primaryClass   = "hep-th",
      SLACcitation   = "%%CITATION = ARXIV:1602.01102;%%"
}

@article{Coleman:1977py,
      author         = "Coleman, Sidney R.",
      title          = "{The Fate of the False Vacuum. 1. Semiclassical Theory}",
      journal        = "Phys. Rev.",
      volume         = "D15",
      year           = "1977",
      pages          = "2929-2936",
      doi            = "10.1103/PhysRevD.15.2929, 10.1103/PhysRevD.16.1248",
      note           = "[Erratum: Phys. Rev.D16,1248(1977)]",
      reportNumber   = "HUTP-77-A004",
      SLACcitation   = "%%CITATION = PHRVA,D15,2929;%%"
}

@article{Callan:1977pt,
      author         = "Callan, Jr., Curtis G. and Coleman, Sidney R.",
      title          = "{The Fate of the False Vacuum. 2. First Quantum
                        Corrections}",
      journal        = "Phys. Rev.",
      volume         = "D16",
      year           = "1977",
      pages          = "1762-1768",
      doi            = "10.1103/PhysRevD.16.1762",
      reportNumber   = "HUTP-77-A032",
      SLACcitation   = "%%CITATION = PHRVA,D16,1762;%%"
}

@article{Coleman:1980aw,
      author         = "Coleman, Sidney R. and De Luccia, Frank",
      title          = "{Gravitational Effects on and of Vacuum Decay}",
      journal        = "Phys. Rev.",
      volume         = "D21",
      year           = "1980",
      pages          = "3305",
      doi            = "10.1103/PhysRevD.21.3305",
      reportNumber   = "SLAC-PUB-2463",
      SLACcitation   = "%%CITATION = PHRVA,D21,3305;%%"
}

@article{Bezrukov:2003er,
      author         = "Bezrukov, F. L. and Levkov, D. and Rebbi, C. and Rubakov,
                        V. A. and Tinyakov, P.",
      title          = "{Semiclassical study of baryon and lepton number
                        violation in high-energy electroweak collisions}",
      journal        = "Phys. Rev.",
      volume         = "D68",
      year           = "2003",
      pages          = "036005",
      doi            = "10.1103/PhysRevD.68.036005",
      eprint         = "hep-ph/0304180",
      archivePrefix  = "arXiv",
      primaryClass   = "hep-ph",
      reportNumber   = "INR-TH-2003-5, BUHEP-03-09",
      SLACcitation   = "%%CITATION = HEP-PH/0304180;%%"
}

@article{Levkov:2004tf,
      author         = "Levkov, D. G. and Sibiryakov, S. M.",
      title          = "{Induced tunneling in QFT: Soliton creation in collisions
                        of highly energetic particles}",
      journal        = "Phys. Rev.",
      volume         = "D71",
      year           = "2005",
      pages          = "025001",
      doi            = "10.1103/PhysRevD.71.025001",
      eprint         = "hep-th/0410198",
      archivePrefix  = "arXiv",
      primaryClass   = "hep-th",
      SLACcitation   = "%%CITATION = HEP-TH/0410198;%%"
}

@article{Callan:1992rs,
      author         = "Callan, Jr., Curtis G. and Giddings, Steven B. and
                        Harvey, Jeffrey A. and Strominger, Andrew",
      title          = "{Evanescent black holes}",
      journal        = "Phys. Rev.",
      volume         = "D45",
      year           = "1992",
      number         = "4",
      pages          = "R1005",
      doi            = "10.1103/PhysRevD.45.R1005",
      eprint         = "hep-th/9111056",
      archivePrefix  = "arXiv",
      primaryClass   = "hep-th",
      reportNumber   = "UCSB-TH-91-54, EFI-91-67, PUPT-1294",
      SLACcitation   = "%%CITATION = HEP-TH/9111056;%%"
}

@article{Hartle:1976tp,
      author         = "Hartle, J. B. and Hawking, S. W.",
      title          = "{Path Integral Derivation of Black Hole Radiance}",
      journal        = "Phys. Rev.",
      volume         = "D13",
      year           = "1976",
      pages          = "2188-2203",
      doi            = "10.1103/PhysRevD.13.2188",
      SLACcitation   = "%%CITATION = PHRVA,D13,2188;%%"
}

@article{Unruh:1976db,
      author         = "Unruh, W. G.",
      title          = "{Notes on black hole evaporation}",
      journal        = "Phys. Rev.",
      volume         = "D14",
      year           = "1976",
      pages          = "870",
      doi            = "10.1103/PhysRevD.14.870",
      SLACcitation   = "%%CITATION = PHRVA,D14,870;%%"
}

@article{Berezin:1987ea,
      author         = "Berezin, V. A. and Kuzmin, V. A. and Tkachev, I. I.",
      title          = "{O(3) Invariant Tunneling in General Relativity}",
      journal        = "Phys. Lett.",
      volume         = "B207",
      year           = "1988",
      pages          = "397-403",
      doi            = "10.1016/0370-2693(88)90672-7",
      reportNumber   = "NBI-HE-87-85",
      SLACcitation   = "%%CITATION = PHLTA,B207,397;%%"
}

@article{Arnold:1989cq,
      author         = "Arnold, Peter Brockway",
      title          = "{Gravity and false vacuum decay rates: O(3) solutions}",
      journal        = "Nucl. Phys.",
      volume         = "B346",
      year           = "1990",
      pages          = "160-192",
      doi            = "10.1016/0550-3213(90)90243-7",
      reportNumber   = "ANL-HEP-PR-89-123",
      SLACcitation   = "%%CITATION = NUPHA,B346,160;%%"
}

@article{Berezin:1990qs,
      author         = "Berezin, V. A. and Kuzmin, V. A. and Tkachev, I. I.",
      title          = "{Black holes initiate false vacuum decay}",
      journal        = "Phys. Rev.",
      volume         = "D43",
      year           = "1991",
      pages          = "3112-3116",
      doi            = "10.1103/PhysRevD.43.R3112",
      reportNumber   = "UCLA-90-TEP-71",
      SLACcitation   = "%%CITATION = PHRVA,D43,3112;%%"
}

@article{Gorbunov:2017fhq,
      author         = "Gorbunov, Dmitry and Levkov, Dmitry and Panin, Alexander",
      title          = "{Fatal youth of the Universe: black hole threat for the
                        electroweak vacuum during preheating}",
      journal        = "JCAP",
      volume         = "1710",
      year           = "2017",
      number         = "10",
      pages          = "016",
      doi            = "10.1088/1475-7516/2017/10/016",
      eprint         = "1704.05399",
      archivePrefix  = "arXiv",
      primaryClass   = "astro-ph.CO",
      reportNumber   = "INR-TH-2017-008",
      SLACcitation   = "%%CITATION = ARXIV:1704.05399;%%"
}

@article{Kohri:2017ybt,
      author         = "Kohri, Kazunori and Matsui, Hiroki",
      title          = "{Electroweak Vacuum Collapse induced by Vacuum
                        Fluctuations of the Higgs Field around Evaporating Black
                        Holes}",
      journal        = "Phys. Rev.",
      volume         = "D98",
      year           = "2018",
      number         = "12",
      pages          = "123509",
      doi            = "10.1103/PhysRevD.98.123509",
      eprint         = "1708.02138",
      archivePrefix  = "arXiv",
      primaryClass   = "hep-ph",
      reportNumber   = "KEK-TH-1989, KEK-Cosmo-208, KEK-COSMO-208",
      SLACcitation   = "%%CITATION = ARXIV:1708.02138;%%"
}

@article{Hiscock:1987hn,
      author         = "Hiscock, W. A.",
      title          = "{Can black holes nucleate vacuum phase transitions?}",
      journal        = "Phys. Rev.",
      volume         = "D35",
      year           = "1987",
      pages          = "1161-1170",
      doi            = "10.1103/PhysRevD.35.1161",
      SLACcitation   = "%%CITATION = PHRVA,D35,1161;%%"
}

@article{Gregory:2013hja,
      author         = "Gregory, Ruth and Moss, Ian G. and Withers, Benjamin",
      title          = "{Black holes as bubble nucleation sites}",
      journal        = "JHEP",
      volume         = "03",
      year           = "2014",
      pages          = "081",
      doi            = "10.1007/JHEP03(2014)081",
      eprint         = "1401.0017",
      archivePrefix  = "arXiv",
      primaryClass   = "hep-th",
      reportNumber   = "DCPT-13-43",
      SLACcitation   = "%%CITATION = ARXIV:1401.0017;%%"
}

@article{Burda:2015isa,
      author         = "Burda, Philipp and Gregory, Ruth and Moss, Ian",
      title          = "{Gravity and the stability of the Higgs vacuum}",
      journal        = "Phys. Rev. Lett.",
      volume         = "115",
      year           = "2015",
      pages          = "071303",
      doi            = "10.1103/PhysRevLett.115.071303",
      eprint         = "1501.04937",
      archivePrefix  = "arXiv",
      primaryClass   = "hep-th",
      reportNumber   = "DCPT-15-03",
      SLACcitation   = "%%CITATION = ARXIV:1501.04937;%%"
}

@article{Burda:2015yfa,
      author         = "Burda, Philipp and Gregory, Ruth and Moss, Ian",
      title          = "{Vacuum metastability with black holes}",
      journal        = "JHEP",
      volume         = "08",
      year           = "2015",
      pages          = "114",
      doi            = "10.1007/JHEP08(2015)114",
      eprint         = "1503.07331",
      archivePrefix  = "arXiv",
      primaryClass   = "hep-th",
      reportNumber   = "DCPT-15-11",
      SLACcitation   = "%%CITATION = ARXIV:1503.07331;%%"
}

@article{Burda:2016mou,
      author         = "Burda, Philipp and Gregory, Ruth and Moss, Ian",
      title          = "{The fate of the Higgs vacuum}",
      journal        = "JHEP",
      volume         = "06",
      year           = "2016",
      pages          = "025",
      doi            = "10.1007/JHEP06(2016)025",
      eprint         = "1601.02152",
      archivePrefix  = "arXiv",
      primaryClass   = "hep-th",
      reportNumber   = "DCPT-16-01",
      SLACcitation   = "%%CITATION = ARXIV:1601.02152;%%"
}

@article{Mukaida:2017bgd,
      author         = "Mukaida, Kyohei and Yamada, Masaki",
      title          = "{False Vacuum Decay Catalyzed by Black Holes}",
      journal        = "Phys. Rev.",
      volume         = "D96",
      year           = "2017",
      number         = "10",
      pages          = "103514",
      doi            = "10.1103/PhysRevD.96.103514",
      eprint         = "1706.04523",
      archivePrefix  = "arXiv",
      primaryClass   = "hep-th",
      reportNumber   = "IPMU-17-0069",
      SLACcitation   = "%%CITATION = ARXIV:1706.04523;%%"
}

@article{1205.2893,
      author         = "Bezrukov, Fedor and Kalmykov, Mikhail {\relax Yu}. and
                        Kniehl, Bernd A. and Shaposhnikov, Mikhail",
      title          = "{Higgs Boson Mass and New Physics}",
      booktitle      = "{Helmholtz Alliance Linear Collider Forum: Proceedings of
                        the Workshops Hamburg, Munich, Hamburg 2010-2012,
                        Germany}",
      journal        = "JHEP",
      volume         = "10",
      year           = "2012",
      pages          = "140",
      doi            = "10.1007/JHEP10(2012)140",
      eprint         = "1205.2893",
      archivePrefix  = "arXiv",
      primaryClass   = "hep-ph",
      reportNumber   = "DESY-12-074",
      SLACcitation   = "%%CITATION = ARXIV:1205.2893;%%"
}

@article{1205.6497,
      author         = "Degrassi, Giuseppe and Di Vita, Stefano and Elias-Miro,
                        Joan and Espinosa, Jose R. and Giudice, Gian F. and
                        Isidori, Gino and Strumia, Alessandro",
      title          = "{Higgs mass and vacuum stability in the Standard Model at
                        NNLO}",
      journal        = "JHEP",
      volume         = "08",
      year           = "2012",
      pages          = "098",
      doi            = "10.1007/JHEP08(2012)098",
      eprint         = "1205.6497",
      archivePrefix  = "arXiv",
      primaryClass   = "hep-ph",
      reportNumber   = "CERN-PH-TH-2012-134, RM3-TH-12-9",
      SLACcitation   = "%%CITATION = ARXIV:1205.6497;%%"
}

@article{1307.3536,
      author         = "Buttazzo, Dario and Degrassi, Giuseppe and Giardino, Pier
                        Paolo and Giudice, Gian F. and Sala, Filippo and Salvio,
                        Alberto and Strumia, Alessandro",
      title          = "{Investigating the near-criticality of the Higgs boson}",
      journal        = "JHEP",
      volume         = "12",
      year           = "2013",
      pages          = "089",
      doi            = "10.1007/JHEP12(2013)089",
      eprint         = "1307.3536",
      archivePrefix  = "arXiv",
      primaryClass   = "hep-ph",
      reportNumber   = "CERN-PH-TH-2013-166, FTUAM-13-20, IFT-UAM-CSIC-13-081,
                        IFUP-TH",
      SLACcitation   = "%%CITATION = ARXIV:1307.3536;%%"
}

@article{Coleman:1977th,
      author         = "Coleman, Sidney R. and Glaser, V. and Martin, Andre",
      title          = "{Action Minima Among Solutions to a Class of Euclidean
                        Scalar Field Equations}",
      journal        = "Commun. Math. Phys.",
      volume         = "58",
      year           = "1978",
      pages          = "211-221",
      doi            = "10.1007/BF01609421",
      reportNumber   = "CERN-TH-2364",
      SLACcitation   = "%%CITATION = CMPHA,58,211;%%"
}

@article{Blum:2016ipp,
      author         = "Blum, Kfir and Honda, Masazumi and Sato, Ryosuke and
                        Takimoto, Masahiro and Tobioka, Kohsaku",
      title          = "{O($N$) Invariance of the Multi-Field Bounce}",
      journal        = "JHEP",
      volume         = "05",
      year           = "2017",
      pages          = "109",
      doi            = "10.1007/JHEP05(2017)109, 10.1007/JHEP06(2017)060",
      note           = "[Erratum: JHEP06,060(2017)]",
      eprint         = "1611.04570",
      archivePrefix  = "arXiv",
      primaryClass   = "hep-th",
      SLACcitation   = "%%CITATION = ARXIV:1611.04570;%%"
}

@article{Tetradis:2016vqb,
      author         = "Tetradis, Nikolaos",
      title          = "{Black holes and Higgs stability}",
      journal        = "JCAP",
      volume         = "1609",
      year           = "2016",
      number         = "09",
      pages          = "036",
      doi            = "10.1088/1475-7516/2016/09/036",
      eprint         = "1606.04018",
      archivePrefix  = "arXiv",
      primaryClass   = "hep-ph",
      SLACcitation   = "%%CITATION = ARXIV:1606.04018;%%"
}

@article{Canko:2017ebb,
      author         = "Canko, D. and Gialamas, I. and Jelic-Cizmek, G. and
                        Riotto, A. and Tetradis, N.",
      title          = "{On the Catalysis of the Electroweak Vacuum Decay by
                        Black Holes at High Temperature}",
      journal        = "Eur. Phys. J.",
      volume         = "C78",
      year           = "2018",
      number         = "4",
      pages          = "328",
      doi            = "10.1140/epjc/s10052-018-5808-y",
      eprint         = "1706.01364",
      archivePrefix  = "arXiv",
      primaryClass   = "hep-th",
      SLACcitation   = "%%CITATION = ARXIV:1706.01364;%%"
}

@article{Levkov:2004ij,
      author         = "Levkov, D. and Sibiryakov, S.",
      title          = "{Real-time instantons and suppression of
                        collision-induced tunneling}",
      journal        = "JETP Lett.",
      volume         = "81",
      year           = "2005",
      pages          = "53-57",
      doi            = "10.1134/1.1887914",
      note           = "[Pisma Zh. Eksp. Teor. Fiz.81,60(2005)]",
      eprint         = "hep-th/0412253",
      archivePrefix  = "arXiv",
      primaryClass   = "hep-th",
      SLACcitation   = "%%CITATION = HEP-TH/0412253;%%"
}

@article{Candelas:1980zt,
    author = "Candelas, P.",
    doi = "10.1103/PhysRevD.21.2185",
    journal = "Phys.\ Rev.\ D",
    pages = "2185--2202",
    title = "{Vacuum Polarization in Schwarzschild Space-Time}",
    volume = "21",
    year = "1980"
}

@article{Rubakov:1992ec,
    author = "Rubakov, V. A. and Son, D. T. and Tinyakov, P. G.",
    title = "{Classical boundary value problem for instanton transitions at high-energies}",
    doi = "10.1016/0370-2693(92)90994-F",
    journal = "Phys. Lett. B",
    volume = "287",
    pages = "342--348",
    year = "1992"
}

@article{Demidov:2011dk,
    author = "Demidov, S. V. and Levkov, D. G.",
    title = "{Soliton-antisoliton pair production in particle collisions}",
    eprint = "1103.0013",
    archivePrefix = "arXiv",
    primaryClass = "hep-th",
    doi = "10.1103/PhysRevLett.107.071601",
    journal = "Phys. Rev. Lett.",
    volume = "107",
    pages = "071601",
    year = "2011"
}

@article{Demidov:2015nea,
    author = "Demidov, S. V. and Levkov, D. G.",
    title = "{Semiclassical description of soliton-antisoliton pair production in particle collisions}",
    eprint = "1509.07125",
    archivePrefix = "arXiv",
    primaryClass = "hep-th",
    reportNumber = "INR-TH-2015-023",
    doi = "10.1007/JHEP11(2015)066",
    journal = "JHEP",
    volume = "11",
    pages = "066",
    year = "2015"
}

@article{Bezrukov:2003tg,
    author = "Bezrukov, F. L. and Levkov, D.",
    title = "{Dynamical tunneling of bound systems through a potential barrier: complex way to the top}",
    eprint = "quant-ph/0312144",
    archivePrefix = "arXiv",
    doi = "10.1134/1.1757681",
    journal = "J. Exp. Theor. Phys.",
    volume = "98",
    pages = "820--836",
    year = "2004"
}

@article{Levkov:2007yn,
    author = "Levkov, D. G. and Panin, A. G. and Sibiryakov, S. M.",
    title = "{Unstable Semiclassical Trajectories in Tunneling}",
    eprint = "0707.0433",
    archivePrefix = "arXiv",
    primaryClass = "quant-ph",
    reportNumber = "CERN-PH-TH-2007-109",
    doi = "10.1103/PhysRevLett.99.170407",
    journal = "Phys. Rev. Lett.",
    volume = "99",
    pages = "170407",
    year = "2007"
}

@article{Demidov:2015bua,
    author = "Demidov, Sergei and Levkov, Dmitry",
    title = "{High-energy limit of collision-induced false vacuum decay}",
    eprint = "1503.06339",
    archivePrefix = "arXiv",
    primaryClass = "hep-ph",
    reportNumber = "INR-TH-2015-009",
    doi = "10.1007/JHEP06(2015)123",
    journal = "JHEP",
    volume = "06",
    pages = "123",
    year = "2015"
}

@article{Coleman:1978ae,
    author = "Coleman, Sidney R.",
    editor = "Shifman, Mikhail A.",
    title = "{The Uses of Instantons}",
    reportNumber = "HUTP-78-A004",
    journal = "Subnucl. Ser.",
    volume = "15",
    pages = "805",
    year = "1979"
}

@ARTICLE{0806.0299,
       author = {{Byeon}, Jaeyoung and {Jeanjean}, Louis and {Mari{\c{s}}}, Mihai}, 
        title = "{Symmetry and monotonicity of least energy solutions}",
      journal = {arXiv e-prints},
         year = 2008,
        month = jun,
          eids = {arXiv:0806.0299},
        pagess = {arXiv:0806.0299},
archivePrefix = {arXiv},
       eprint = {0806.0299},
 primaryClass = {math.AP},
       adsurl = {https://ui.adsabs.harvard.edu/abs/2008arXiv0806.0299B}
}

@article{Takahashi_2003,
	doi = {10.1088/0305-4470/36/29/305},
	url = {https://doi.org/10.1088/0305-4470/36/29/305},
	year = 2003,
	month = {jul},
	publisher = {{IOP} Publishing},
	volume = {36},
	number = {29},
	pages = {7953--7987},
	author = {Kin Takahashi and Kensuke S Ikeda},
	title = {Complex-classical mechanism of the tunnelling process in strongly coupled 1.5-dimensional barrier systems},
	journal = {Journal of Physics A: Mathematical and General}
}

@article{Takahashi_2005,
	doi = {10.1209/epl/i2004-10538-1},
	url = {https://doi.org/10.1209/epl/i2004-10538-1},
	year = 2005,
	month = {jul},
	publisher = {{IOP} Publishing},
	volume = {71},
	number = {2},
	pages = {193--199},
	author = {K Takahashi and K. S Ikeda},
	title = {An intrinsic multi-dimensional mechanism of barrier tunneling},
	journal = {Europhysics Letters ({EPL})}
}

@article{Levkov:2008csa,
    author = "Levkov, D. G. and Panin, A. G. and Sibiryakov, S. M.",
    title = "{Signatures of unstable semiclassical trajectories in tunneling}",
    eprint = "0811.3391",
    archivePrefix = "arXiv",
    primaryClass = "quant-ph",
    doi = "10.1088/1751-8113/42/20/205102",
    journal = "J. Phys. A",
    volume = "42",
    pages = "205102",
    year = "2009"
}

@article{Hayashi:2020ocn,
    author = "Hayashi, Takumi and Kamada, Kohei and Oshita, Naritaka and Yokoyama, Jun'ichi",
    title = "{On catalyzed vacuum decay around a radiating black hole and the crisis of the electroweak vacuum}",
    eprint = "2005.12808",
    archivePrefix = "arXiv",
    primaryClass = "hep-th",
    reportNumber = "RESCEU-10/20",
    doi = "10.1007/JHEP08(2020)088",
    journal = "JHEP",
    volume = "08",
    pages = "088",
    year = "2020"
}

@article{Cherman:2014sba,
    author = "Cherman, Aleksey and Unsal, Mithat",
    title = "{Real-Time Feynman Path Integral Realization of Instantons}",
    eprint = "1408.0012",
    archivePrefix = "arXiv",
    primaryClass = "hep-th",
    reportNumber = "FTPI-MINN-14-18, UMN-TH-3343-14",
    month = "7",
    year = "2014"
}

@article{Turok:2013dfa,
    author = "Turok, Neil",
    title = "{On Quantum Tunneling in Real Time}",
    eprint = "1312.1772",
    archivePrefix = "arXiv",
    primaryClass = "quant-ph",
    doi = "10.1088/1367-2630/16/6/063006",
    journal = "New J. Phys.",
    volume = "16",
    pages = "063006",
    year = "2014"
}

@article{Bramberger:2016yog,
    author = "Bramberger, Sebastian F. and Lavrelashvili, George and Lehners, Jean-Luc",
    title = "{Quantum tunneling from paths in complex time}",
    eprint = "1605.02751",
    archivePrefix = "arXiv",
    primaryClass = "hep-th",
    doi = "10.1103/PhysRevD.94.064032",
    journal = "Phys. Rev. D",
    volume = "94",
    number = "6",
    pages = "064032",
    year = "2016"
}

@article{GarciaBellido:1996qt,
    author = "Garcia-Bellido, Juan and Linde, Andrei D. and Wands, David",
    title = "{Density perturbations and black hole formation in hybrid inflation}",
    eprint = "astro-ph/9605094",
    archivePrefix = "arXiv",
    reportNumber = "SU-ITP-96-20, SUSSEX-AST-96-5-1",
    doi = "10.1103/PhysRevD.54.6040",
    journal = "Phys. Rev. D",
    volume = "54",
    pages = "6040--6058",
    year = "1996"
}

@article{Sher:1988mj,
    author = "Sher, Marc",
    title = "{Electroweak Higgs Potentials and Vacuum Stability}",
    reportNumber = "WU-TH-88-8",
    doi = "10.1016/0370-1573(89)90061-6",
    journal = "Phys. Rept.",
    volume = "179",
    pages = "273--418",
    year = "1989"
}

@article{Flores:1982rv,
    author = "Flores, Ricardo A. and Sher, Marc",
    title = "{Upper Limits to Fermion Masses in the {Glashow-Weinberg-Salam} Model}",
    reportNumber = "UCI-82-89",
    doi = "10.1103/PhysRevD.27.1679",
    journal = "Phys. Rev. D",
    volume = "27",
    pages = "1679",
    year = "1983"
}

@article{Isidori:2001bm,
    author = "Isidori, Gino and Ridolfi, Giovanni and Strumia, Alessandro",
    title = "{On the metastability of the standard model vacuum}",
    eprint = "hep-ph/0104016",
    archivePrefix = "arXiv",
    reportNumber = "CERN-TH-2001-092, LNF-01-014-P, GEF-TH-6-01, IFUP-TH-2001-11",
    doi = "10.1016/S0550-3213(01)00302-9",
    journal = "Nucl. Phys. B",
    volume = "609",
    pages = "387--409",
    year = "2001"
}

@article{Bednyakov:2015sca,
    author = "Bednyakov, A. V. and Kniehl, B. A. and Pikelner, A. F. and Veretin, O. L.",
    title = "{Stability of the Electroweak Vacuum: Gauge Independence and Advanced Precision}",
    eprint = "1507.08833",
    archivePrefix = "arXiv",
    primaryClass = "hep-ph",
    reportNumber = "DESY-15-131",
    doi = "10.1103/PhysRevLett.115.201802",
    journal = "Phys. Rev. Lett.",
    volume = "115",
    number = "20",
    pages = "201802",
    year = "2015"
}

@article{Fujita:2014hha,
    author = "Fujita, Tomohiro and Kawasaki, Masahiro and Harigaya, Keisuke and Matsuda, Ryo",
    title = "{Baryon asymmetry, dark matter, and density perturbation from primordial black holes}",
    eprint = "1401.1909",
    archivePrefix = "arXiv",
    primaryClass = "astro-ph.CO",
    reportNumber = "IPMU-14-0009, ICRR-REPORT-668-2013-17",
    doi = "10.1103/PhysRevD.89.103501",
    journal = "Phys. Rev. D",
    volume = "89",
    number = "10",
    pages = "103501",
    year = "2014"
}

@article{Allahverdi:2017sks,
    author = "Allahverdi, Rouzbeh and Dent, James and Osinski, Jacek",
    title = "{Nonthermal production of dark matter from primordial black holes}",
    eprint = "1711.10511",
    archivePrefix = "arXiv",
    primaryClass = "astro-ph.CO",
    doi = "10.1103/PhysRevD.97.055013",
    journal = "Phys. Rev. D",
    volume = "97",
    number = "5",
    pages = "055013",
    year = "2018"
}

@article{Lennon:2017tqq,
    author = "Lennon, Olivier and March-Russell, John and Petrossian-Byrne, Rudin and Tillim, Hannah",
    title = "{Black Hole Genesis of Dark Matter}",
    eprint = "1712.07664",
    archivePrefix = "arXiv",
    primaryClass = "hep-ph",
    doi = "10.1088/1475-7516/2018/04/009",
    journal = "JCAP",
    volume = "04",
    pages = "009",
    year = "2018"
}

@article{Morrison:2018xla,
    author = "Morrison, Logan and Profumo, Stefano and Yu, Yan",
    title = "{Melanopogenesis: Dark Matter of (almost) any Mass and Baryonic Matter from the Evaporation of Primordial Black Holes weighing a Ton (or less)}",
    eprint = "1812.10606",
    archivePrefix = "arXiv",
    primaryClass = "astro-ph.CO",
    doi = "10.1088/1475-7516/2019/05/005",
    journal = "JCAP",
    volume = "05",
    pages = "005",
    year = "2019"
}

@article{Hooper:2019gtx,
    author = "Hooper, Dan and Krnjaic, Gordan and McDermott, Samuel D.",
    title = "{Dark Radiation and Superheavy Dark Matter from Black Hole Domination}",
    eprint = "1905.01301",
    archivePrefix = "arXiv",
    primaryClass = "hep-ph",
    reportNumber = "FERMILAB-PUB-19-186-A",
    doi = "10.1007/JHEP08(2019)001",
    journal = "JHEP",
    volume = "08",
    pages = "001",
    year = "2019"
}

@article{Hooper:2020evu,
    author = "Hooper, Dan and Krnjaic, Gordan and March-Russell, John and McDermott, Samuel D. and Petrossian-Byrne, Rudin",
    title = "{Hot Gravitons and Gravitational Waves From Kerr Black Holes in the Early Universe}",
    eprint = "2004.00618",
    archivePrefix = "arXiv",
    primaryClass = "astro-ph.CO",
    reportNumber = "FERMILAB-PUB-20-125-A-T",
    month = "4",
    year = "2020"
}

@book{doi:10.1142/3768,
author = {Tomsovic, Steven},
title = {Tunneling in Complex Systems},
publisher = {WORLD SCIENTIFIC},
year = {1998},
doi = {https://doi.org/10.1142/3768},
address = {},
edition   = {},
URL = {https://www.worldscientific.com/doi/abs/10.1142/3768}
}

@article{Bonini:1999kj,
    author = "Bonini, G. F. and Cohen, Andrew G. and Rebbi, C. and Rubakov, V. A.",
    title = "{The Semiclassical description of tunneling in scattering with multiple degrees of freedom}",
    eprint = "hep-ph/9901226",
    archivePrefix = "arXiv",
    reportNumber = "BUHEP-99-1, BU-HEP-99-1",
    doi = "10.1103/PhysRevD.60.076004",
    journal = "Phys. Rev. D",
    volume = "60",
    pages = "076004",
    year = "1999"
}

@incollection{Miller,
author = {Miller, William H.},
publisher = {John Wiley \& Sons, Ltd},
isbn = {9780470143773},
title = {Classical-Limit Quantum Mechanics and the Theory of Molecular Collisions},
booktitle = {Advances in Chemical Physics},
chapter = {},
pages = {69-177},
doi = {https://doi.org/10.1002/9780470143773.ch2},
url = {https://onlinelibrary.wiley.com/doi/abs/10.1002/9780470143773.ch2},
year = {1974}
}

@article{Linde:1981zj,
    author = "Linde, Andrei D.",
    title = "{Decay of the False Vacuum at Finite Temperature}",
    reportNumber = "LEBEDEV-81-265",
    doi = "10.1016/0550-3213(83)90072-X",
    journal = "Nucl. Phys. B",
    volume = "216",
    pages = "421",
    year = "1983",
    note = "[Erratum: Nucl.Phys.B 223, 544 (1983)]"
}

@article{Markkanen:2018pdo,
    author = "Markkanen, Tommi and Rajantie, Arttu and Stopyra, Stephen",
    title = "{Cosmological Aspects of Higgs Vacuum Metastability}",
    eprint = "1809.06923",
    archivePrefix = "arXiv",
    primaryClass = "astro-ph.CO",
    reportNumber = "IMPERIAL/TP/2018/TM/05",
    doi = "10.3389/fspas.2018.00040",
    journal = "Front. Astron. Space Sci.",
    volume = "5",
    pages = "40",
    year = "2018"
}

@article{Miyachi:2021bwd,
    author = "Miyachi, Taiga and Soda, Jiro",
    title = "{False vacuum decay in a two-dimensional black hole spacetime}",
    eprint = "2102.02462",
    archivePrefix = "arXiv",
    primaryClass = "gr-qc",
    reportNumber = "KOBE-COSMO-21-02",
    doi = "10.1103/PhysRevD.103.085009",
    journal = "Phys. Rev. D",
    volume = "103",
    number = "8",
    pages = "085009",
    year = "2021"
}

@article{Dai:2019eei,
    author = "Dai, De-Chang and Gregory, Ruth and Stojkovic, Dejan",
    title = "{Connecting the Higgs Potential and Primordial Black Holes}",
    eprint = "1909.00773",
    archivePrefix = "arXiv",
    primaryClass = "hep-ph",
    reportNumber = "DCPT-19/25",
    doi = "10.1103/PhysRevD.101.125012",
    journal = "Phys. Rev. D",
    volume = "101",
    number = "12",
    pages = "125012",
    year = "2020"
}

@article{Dong:2015yjs,
    author = "Dong, Ruifeng and Kinney, William H and Stojkovic, Dejan",
    title = "{Gravitational wave production by Hawking radiation from rotating primordial black holes}",
    eprint = "1511.05642",
    archivePrefix = "arXiv",
    primaryClass = "astro-ph.CO",
    doi = "10.1088/1475-7516/2016/10/034",
    journal = "JCAP",
    volume = "10",
    pages = "034",
    year = "2016"
}

@article{Shkerin:2021rhy,
	author = "Shkerin, Andrey and Sibiryakov, Sergey",
	title = "{Black hole induced false vacuum decay: the role of greybody factors}",
	eprint = "2111.08017",
	archivePrefix = "arXiv",
	primaryClass = "hep-th",
	reportNumber = "FTPI-MINN-21-20, UMN-TH-4103/2, INR-TH-2021-021",
	doi = "10.1007/JHEP08(2022)161",
	journal = "JHEP",
	volume = "08",
	pages = "161",
	year = "2022"
}

@article{Rossi:2025fix,
    author = "Rossi, Giuseppe",
    title = "{Vacuum Decay around Black Holes formed from Collapse}",
    eprint = "2512.23048",
    archivePrefix = "arXiv",
    primaryClass = "hep-th",
    month = "12",
    year = "2025"
}

@article{He:2024wvt,
    author = "He, Minxi and Kohri, Kazunori and Mukaida, Kyohei and Yamada, Masaki",
    title = "{Thermalization and hotspot formation around small primordial black holes}",
    eprint = "2407.15926",
    archivePrefix = "arXiv",
    primaryClass = "hep-ph",
    reportNumber = "KEK-TH-2639, EK-TH-2639, TU-1238, CTPU-PTC-24-22, KEK-Cosmo-0351,
  KEK-QUP-2024-0018",
    doi = "10.1088/1475-7516/2024/10/080",
    journal = "JCAP",
    volume = "10",
    pages = "080",
    year = "2024"
}

@article{Hamaide:2023ayu,
    author = "Hamaide, Louis and Heurtier, Lucien and Hu, Shi-Qian and Cheek, Andrew",
    title = "{Primordial black holes are true vacuum nurseries}",
    eprint = "2311.01869",
    archivePrefix = "arXiv",
    primaryClass = "hep-ph",
    doi = "10.1016/j.physletb.2024.138895",
    journal = "Phys. Lett. B",
    volume = "856",
    pages = "138895",
    year = "2024"
}

@article{Strumia:2022jil,
    author = "Strumia, Alessandro",
    title = "{Black holes don{\textquoteright}t source fast Higgs vacuum decay}",
    eprint = "2209.05504",
    archivePrefix = "arXiv",
    primaryClass = "hep-ph",
    doi = "10.1007/JHEP03(2023)039",
    journal = "JHEP",
    volume = "03",
    pages = "039",
    year = "2023"
}

@article{Hayashi:2021kro,
    author = "Hayashi, Takumi and Kamada, Kohei and Oshita, Naritaka and Yokoyama, Jun'ichi",
    title = "{Vacuum decay in the Lorentzian path integral}",
    eprint = "2112.09284",
    archivePrefix = "arXiv",
    primaryClass = "hep-th",
    reportNumber = "RESCEU-24/21, RIKEN-iTHEMS-Report-21",
    doi = "10.1088/1475-7516/2022/05/041",
    journal = "JCAP",
    volume = "05",
    number = "05",
    pages = "041",
    year = "2022"
}

@article{Briaud:2022few,
    author = "Briaud, Vadim and Shkerin, Andrey and Sibiryakov, Sergey",
    title = "{Thermal false vacuum decay around black holes}",
    eprint = "2210.08028",
    archivePrefix = "arXiv",
    primaryClass = "gr-qc",
    reportNumber = "FTPI-MINN-22-28, UMN-TH-4203/22",
    doi = "10.1103/PhysRevD.106.125001",
    journal = "Phys. Rev. D",
    volume = "106",
    number = "12",
    pages = "125001",
    year = "2022"
}

@article{Carr:2020gox,
    author = "Carr, Bernard and Kohri, Kazunori and Sendouda, Yuuiti and Yokoyama, Jun'ichi",
    title = "{Constraints on primordial black holes}",
    eprint = "2002.12778",
    archivePrefix = "arXiv",
    primaryClass = "astro-ph.CO",
    reportNumber = "RESCEU-03/20; KEK-Cosmo-249; KEK-TH-2199; IPMU20-0024",
    doi = "10.1088/1361-6633/ac1e31",
    journal = "Rept. Prog. Phys.",
    volume = "84",
    number = "11",
    pages = "116902",
    year = "2021"
}

@article{Shkerin:2021zbf,
    author = "Shkerin, Andrey and Sibiryakov, Sergey",
    title = "{Black hole induced false vacuum decay from first principles}",
    eprint = "2105.09331",
    archivePrefix = "arXiv",
    primaryClass = "hep-th",
    reportNumber = "FTPI-MINN-21-07, UMN-TH-4014/21, INR-TH-2021-011",
    doi = "10.1007/JHEP11(2021)197",
    journal = "JHEP",
    volume = "11",
    pages = "197",
    year = "2021"
}

\end{document}